\documentclass[prd,twocolumn,amsmath,amssymb]{revtex4-1}

\usepackage{dcolumn}
\usepackage{bm}
\usepackage{epsfig}
\usepackage{graphicx}
\usepackage{amsmath}
\usepackage{amssymb}
\usepackage{mathrsfs}
\usepackage{color}
\usepackage[usenames,dvipsnames]{xcolor}
\usepackage{hyperref}
\usepackage[caption=false]{subfig}

\usepackage{amsfonts}

\def\be{\begin{equation}}
\def\ee{\end{equation}}
\def\bea{\begin{eqnarray}}
\def\eea{\end{eqnarray}}
\def\nn{\nonumber}

\def\slashchar#1{\setbox0=\hbox{$#1$}           
   \dimen0=\wd0                                 
   \setbox1=\hbox{/} \dimen1=\wd1               
   \ifdim\dimen0>\dimen1                        
      \rlap{\hbox to \dimen0{\hfil/\hfil}}      
      #1                                        
   \else                                        
      \rlap{\hbox to \dimen1{\hfil$#1$\hfil}}   
      /                                         
   \fi}

\begin{document}

\title{$T_{cc}^+$ decays: Differential spectra and two-body final states}


\author{Sean Fleming}
\email{spf@email.arizona.edu}
\affiliation{Department of Physics and Astronomy,\\
         University of Arizona, Tucson, AZ\ 85721, USA\\}

\author{Reed Hodges}
\email{reed.hodges@duke.edu}
\affiliation{Department of Physics,\\
         Duke University, Durham, NC\ 27708, USA\\}

\author{Thomas Mehen}
\email{mehen@phy.duke.edu}
\affiliation{Department of Physics,\\
         Duke University, Durham, NC\ 27708, USA\\}

\begin{abstract} 
The recently discovered tetraquark, $T_{cc}^+$, has  quark content $cc\bar{u}\bar{d}$ and a 
mass that lies just below open charm thresholds. Hence it is reasonable to expect the state to have a significant molecular component. 
We calculate the decay of the $T_{cc}^+$ in a molecular interpretation using effective field theory. In addition we calculate differential spectra
as a function of the invariant mass of the final state charm meson pair.  These are in good agreement with spectra measured by LHCb. We also point out that if shallow bound states of two pseudoscalar charm mesons exist, then two-body decays to those bound states and a single pion or photon can significantly enhance the width of the $T_{cc}^+$. 

\end{abstract}

\maketitle
 
The LHCb experiment has recently observed a narrow resonance, $T_{cc}^+$, in the final state $D^0 D^0 \pi^+$~\cite{LHCb:2021vvq,LHCb:2021auc} (previously announced in Refs.~\cite{Muheim, Polyakov, An}). They use two different Breit-Wigner line shapes to fit the data.  Using a relativistic $P$-wave two-body Breit-Wigner function with a Blatt-Weisskopf form factor, they find the difference of the resonance mass and the $D^0 D^{*+}$ threshold, $\delta m_{BW}$, and the decay width, $\Gamma_{BW}$, to be~\cite{LHCb:2021vvq,LHCb:2021auc}:
\bea\label{BW_width}
\delta m_{BW} &=& - 273 \pm 61 \pm 5 {}^{+11}_{-14}\, {\rm keV} \, , \nonumber \\   
\Gamma_{BW}  &=&  410 \pm 165 \pm 43 {}^{+18}_{-38}\, {\rm keV} \, .
\eea
When using a unitarized Breit-Wigner profile, LHCb obtains a much smaller width~\cite{LHCb:2021auc}:
\bea\label{BW_width2}
\delta m_{pole} &=& -360\pm40^{+4}_{-0}\, {\rm keV} \, , \nonumber \\   
\Gamma_{pole}  &=&  48 \pm 2_{-14}^{+0}\, {\rm keV} \, .
\eea
The $T_{cc}^+$ is 1.7 MeV below the $D^+D^{*0}$ threshold. LHCb also finds evidence for similar enhancements near the $D^0 D^0$ and $D^0 D^+$ thresholds~\cite{Muheim, Polyakov, An}.
The $T_{cc}^+$  clearly  has the exotic quantum numbers of a tetraquark. The closeness to the $D^0 D^{*+}$ and $D^+ D^{*0}$ thresholds  suggests that the $T_{cc}^+$ could also 
be molecular in character. In this paper we use effective field theory  (EFT)   to calculate the decays $T_{cc}^+ \to D^0 D^0\pi^+, D^+ D^0\pi^0$, and $D^+ D^0 \gamma$ in the molecular scenario. We also consider the possibility of shallow bound states of $D^0 D^0$ and $D^0 D^+$ that we will refer to as $\tilde T_{cc}^0$ and $\tilde T_{cc}^+$, respectively. If these states exist with binding energies of a few MeV then two-body decays $T_{cc}^+\to \tilde T_{cc}^0 \pi^+$ and $T_{cc}^+\to \tilde T_{cc}^+ \pi^0$ can significantly enhance the width of the $T_{cc}^+$. For early work on doubly-heavy tetraquarks, see Refs.~\cite{Dias:2011mi,Janc:2004qn,Zouzou:1986qh,Vijande:2006jf,Brink:1998as,Swanson:2006st,Navarra:2007yw,Du:2012wp,Ebert:2007rn,Manohar:1992nd}, and for more recent theoretical work on the $T_{cc}^+$, see Refs.~\cite{Meng:2021jnw,Agaev:2021vur,Wu:2021kbu,Ling:2021bir, Chen:2021vhg,Dong:2021bvy,Feijoo:2021ppq,Yan:2021wdl,Dai:2021wxi,Weng:2021hje,Huang:2021urd,Chen:2021kad,Xin:2021wcr}.

References~\cite{Meng:2021jnw,Ling:2021bir,Feijoo:2021ppq,Yan:2021wdl} have calculated partial decay widths of $T_{cc}^+$. We will compare their results to ours in more detail below. All calculations of
the widths are small compared to the central value of the width quoted in Eq.~(\ref{BW_width}); however, they are consistent with the width quoted in Eq.~(\ref{BW_width2}). Clearly the extracted width is quite sensitive to how one chooses to fit the data. It is well known that in the presence of multiple scattering channels more complex line shapes are required to fit the data~\cite{Hanhart_2015} and the choice of line shape can have a significant impact on the extracted width. For example, the LHCb experiment measured the  line shape of the $\chi_{c1}(3872)$ in Ref.~\cite{LHCb:2020xds}. When fitting the data with a  Breit-Wigner, they find a width whose central value is 1.39 MeV. This greatly exceeds a bound of $\Gamma[\chi_{c1}(3872)]\leq 131$ keV derived in Ref.~\cite{Mehen:2015efa}. This bound was obtained by extracting $\Gamma[D^{*0}\to D^0\pi^0]$ from the observed width $\Gamma[D^{*+}]$ using isopsin invariance and assuming $\Gamma[\chi_{c1} (3872)\to D^0\bar{D}^0\pi^0]\approx \Gamma[D^{*0}\to D^0\pi^0]$, which is expected on general grounds for a shallow molecular bound state~\cite{Voloshin:2003nt,Fleming:2007rp}. With this theoretical estimate of $\Gamma[\chi_{c1} (3872)\to D^0\bar{D}^0\pi^0]$ one can use the  
branching fraction quoted in the Particle Data Group \cite{Zyla:2020zbs} to obtain a bound on $\Gamma[\chi_{c1}(3872)]$.
Reference~\cite{LHCb:2020xds} also fits the line shape of the $\chi_{c1}(3872)$ with the Flatt\'e line 
shape which properly accounts for two channels (in this case $D^0 \bar{D}^{*0}$ 
and $D^+D^{*-}$). In this case, they find a much narrower lineshape with FWHM (full width half maximum) equal to $0.22^{+0.07+0.11}_{-0.06-0.13}$ MeV. Note that the $D^+D^{*-}$ threshold is 8.2 MeV above the mass of $D^0\bar{D}^{*0}$. The two thresholds in the case of $T_{cc}^+$ are only 1.4 MeV apart so coupled channel effects clearly need to be accounted for. 
 
 The point of this paper is to calculate the strong and electromagnetic decays of $T^+_{cc}$ using an effective theory which treats the constituents of the $T_{cc}^+$ as nonrelativistic particles. We find that our leading order (LO) rates for $T_{cc}^+\to D^0 D^0\pi^+, D^+D^0\pi^0$, and $D^+ D^0\gamma$ are comparable to other recent analyses~\cite{Meng:2021jnw,Ling:2021bir,Feijoo:2021ppq,Yan:2021wdl}. We also calculate differential distributions in the invariant mass, $m_{DD}$, of the charm mesons in the final state. As argued in Ref.~\cite{Dai:2019hrf} for $\chi_{c1}(3872)$,  these distributions are strongly peaked near maximal energy of the pion/photon and are sensitive to the molecular character of $T_{cc}^+$. If $T_{cc}^+$ is a shallow molecule of $D^{0}D^{*+}/D^{+}D^{*0}$, it is conceivable that shallow bound states of $D^0D^+$ and $D^0D^0$ could also exist, opening up another decay channel for $T_{cc}^+$. We calculate these decay rates, which proceed through triangle diagrams in the EFT, under the assumption that the binding energies are between 0 and 5 MeV. If these channels exist they could increase the width of $T_{cc}^+$ by as much as 150 keV.

The effective field theory we will develop is essentially XEFT, first developed in Ref.~\cite{Fleming:2007rp} and further applied in Refs.~\cite{Fleming:2008yn,Fleming:2011xa,Mehen:2011ds,Margaryan:2013tta,Braaten:2010mg,Canham:2009zq,Jansen:2013cba,Jansen:2015lha,Mehen:2015efa,Alhakami:2015uea,Braaten:2015tga,Braaten:2020iye,Braaten:2020nmc,Braaten:2020iqw}; for other EFT analyses of the $\chi_{c1}(3872)$ see Refs.~\cite{AlFiky:2005jd,Baru:2011rs,Valderrama:2012jv,Nieves:2012tt,Baru:2013rta,Guo:2013nza,Guo:2013sya,Baru:2015nea,Schmidt:2018vvl,Sakai:2020ucu,Molina:2020kyu,Sakai:2020crh,Contessi:2020jqa,Wu:2021udi}. The main difference here is that there are two nearly degenerate channels so we will have to solve the coupled channel problem for $D^0 D^{*+}$  and $D^+D^{*0}$. Note in this paper we will only be working at LO, and in this approximation, the predictions can be obtained from effective range theory, as first done for the $\chi_{c1}(3872)$ by Voloshin in Ref.~\cite{Voloshin:2003nt}.
EFT can be used to compute the effect of loops with pions, range corrections, and rescattering effects which are not included in our calculations. Previous experience using XEFT \cite{Dai:2019hrf} will inform our discussion of the uncertainties in the LO calculation 
of this paper. 
 
 The Lagrangian for an effective theory for $T_{cc}^+$ is
 \bea \label{Lag}
 {\mathcal L} &=& H^{* i\dagger}\bigg(i\partial^0+\frac{\nabla^2}{2m_{H^*}}-\delta^*\bigg)H^{* i} \nn \\
&& + H^\dagger\bigg(i\partial^0+\frac{\nabla^2}{2m_H}-\delta\bigg)H  \nn \\
&&+ \frac{g}{f_\pi} H^\dagger \partial^i \pi  H^{*i} + \text{h.c.} \nn \\
&&+ \frac{1}{2}H^\dagger \mu_D \vec{B}^i H^{*i} + \text{h.c.} \nn \\
&&-C_0 (H^{*T}\tau_2 H)^\dagger (H^{*T}\tau_2 H) \nn \\
&&- C_1 (H^{*T}\tau_2 \tau_a H)^\dagger (H^{*T}\tau_2 \tau_a H) \, .
\eea
Here $H$ and $H^{*i}$ are isodoublets of the pseudoscalar and vector fields, 
\bea
H=\left(\begin{array}{c} D^0 \\ D^+ \end{array}\right) \quad H^{*i}=\left(\begin{array}{c} D^{*0i} \\ D^{*+ i} \end{array}\right) \,.
\eea
The first line of Eq.~(\ref{Lag}) contains kinetic terms for charm mesons; $\delta$ and $\delta^*$ are the diagonal matrices of residual masses, whose entries are defined by $\delta^{(*)}_{ii} = M_{D^{(*)i}} - M_{D^0}$, $i=0,+$. 
The second line contains the coupling of charm mesons to pions and photons. The pion field is the matrix
\bea
\pi = \left(\begin{array}{cc} \pi^0/\sqrt{2} & \pi^+ \\ \pi^- & -\pi^0/\sqrt{2}  \end{array}\right) \,,
\eea
$g=0.54$ is the axial coupling of heavy hadron chiral perturbation theory (HH$\chi$PT) \cite{Wise:1992hn,Burdman:1992gh,Yan:1992gz}, $f_\pi = 130$ MeV is the pion decay constant, and the coupling is appropriate for a relativistically normalized pion field. The kinetic terms for pions are not shown since they will not be needed.  In the coupling to the magnetic field, $\vec{B}$,
$\mu_D$ is the matrix of transition magnetic moments, $\mu_D={\rm diag}({\mu_{D^0},\mu_{D^+}})$. This interaction can be derived from HH$\chi$PT~\cite{Amundson:1992yp,Stewart:1998ke}.
We will fix the numerical values of $\mu_{D^+}$ and $\mu_{D^0}$ to give the partial widths $\Gamma[D^{*+}\to D^+\gamma] =1.33$ keV and $\Gamma[D^{*0}\to D^0\gamma]=19.9$ keV at tree-level. The first partial  width is the 
central value that can be obtained from the particle data group (PDG) \cite{Zyla:2020zbs}. The second partial width is obtained by using isospin symmetry to relate $\Gamma[D^{*0}\to D^0 \pi^0]$ to $\Gamma[D^{*+}\to D^+ \pi^0]$ which can be extracted from 
the PDG. Then the branching ratios for $\text{Br}[D^{*0}\to D^{0}\gamma]$ and $\text{Br}[D^{*0}\to D^{0}\pi^0]$ in the PDG can be used to determine $\Gamma[D^{*0}\to D^0\gamma]$~\cite{Hu:2005gf}. Note that in HH$\chi$PT at tree-level 
\bea
\mu_{D^0} = \frac{2e}{3}\beta +\frac{2e}{3m_c} \quad  \mu_{D^+} = -\frac{e}{3}\beta +\frac{2e}{3m_c} \, ,
\eea
where the parameter $\beta$ is defined in Refs.~\cite{Amundson:1992yp,Stewart:1998ke}.
In the heavy quark limit $\mu_{D^0}>0$ and $\mu_{D^+}< 0$; our values for these parameters are consistent with this. 

The terms in the last line of Eq.~(\ref{Lag}) are contact interactions that mediate $DD^{*}$ scattering. The term with $C_0$ mediates scattering in the $I=0$ channel, $D^0 D^{*+} -D^+ D^{*0}$, the term with $C_1$ mediates scattering in the $I=1$ 
channel, $D^0 D^{*+} + D^+ D^{*0}$. The $\tau_a$ are Pauli matrices acting in isospin space. In terms of the charm meson fields, the relevant interactions can be written as
\begin{widetext}
\bea
{\mathcal L} = -\left( \begin{array}{cc} D^{0\,\dagger} D^{*+\,\dagger} & D^{+ \,\dagger} D^{*0\, \dagger}\end{array}\right) 
 \left( \begin{array}{cc} C_0 +C_1& -C_0 +C_1 \\ 
 -C_0 +C_1& C_0 +C_1 \end{array}\right)
 \left(\begin{array}{c} D^0 D^{*+} \\ D^+ D^{*0}\end{array}\right) \, .
 \eea
\end{widetext}
 A similar coupled channel problem appears in, for example, Ref.~\cite{Mehen:2011yh}; solving it yields a T-matrix of the form of Eqs.\,(15)-(17) of that paper. Parameters of that T-matrix must be tuned so that there is a pole in the T-matrix  corresponding to the $T_{cc}^+$. In the vicinity of the pole the T-matrix can be parametrized as
 \bea 
 T= \frac{1}{E+E_T} \left( \begin{array}{cc} g_0^2 & g_0 g_1 \\ g_0 g_1 & g_1^2\end{array} \right) \, .
 \eea
Here $E_T$ is the binding energy of the $T_{cc}^+$, and $g_0$ and $g_1$ give the coupling to each of the channels. The subscript $0(+)$ refers to the charge of the pseudoscalar meson in that channel.
These couplings obey the relation 
\bea \label{wcc}
g_0^2 \Sigma_0^\prime(-E_T) + g_+^2 \Sigma_+^\prime(-E_T)=1 \,,
\eea 
where the derivatives of the self energies are given by
\bea
\Sigma_0^\prime(-E_T) = \frac{\mu_0^2}{2\pi \gamma_0} \quad   \Sigma _+^\prime(-E_T) = \frac{\mu_+^2}{2\pi \gamma_+} \, .
\eea
Here the reduced masses are $\mu_0 = (1/m_{D^0} +1/m_{D^{*+}})^{-1}$, $\mu_+ = (1/m_{D^+} +1/m_{D^{*0}})^{-1}$, and the binding momenta for each channel are given by:
\bea
\gamma_0^2 &=& 2 \mu_0(m_{D_0} +m_{D^{*+}}-m_T) \nn \\
\gamma_+^2 &=& 2 \mu_0(m_{D_+} +m_{D^{*0}}-m_T) \, .
 \eea
The mass of the of $T_{cc}^+$, $m_T$, is taken to be the central value of the Breit-Wigner distribution in Eq.~(\ref{BW_width}).
To satisfy Eq.~(\ref{wcc}) we define 
\bea
g_0^2  = \frac{\cos^2\theta}{\Sigma^\prime_0(-E_T)} \qquad g_+^2  = \frac{\sin^2\theta}{\Sigma^\prime_+(-E_T)} \, ,
\eea
and in what follows we will abbreviate $\sin\theta=s_\theta$ and $\cos\theta=c_\theta$. If the $T_{cc}^+$ is a pure $I=0$ state $g_0 = -g_1$, which is obtained if $\theta= -32.4^\circ$. This is the most likely isospin assignment for the $T_{cc}^+$. In the tetraquark picture of the $T_{cc}^+$ the light $\bar{u}\bar{d}$ are in an $I=0$ diquark configuration and the $I=1$ diquark is expected to be heavier
by $\approx$ 205 MeV (this is the $\Sigma-\Lambda$ mass difference in the light quark as well as bottom and charm quark sectors). 

\begin{table}
\begin{center}
\begin{tabular}{|c||c|c|c|} 
\hline 
& I=0 & I=1 & $\Gamma_{\rm max}$ \\
\hline \hline
$\theta$ & $-32.4^\circ$ & $32.4^\circ$ & $-8.34^\circ$ \\
\hline
$\Gamma[T_{cc}^+\to D^0 D^0 \pi^+]$ & 32 & 32 & 44 \\
\hline
$\Gamma[T_{cc}^+\to D^+ D^0 \pi^0]$ & 15& 3.8& 13 \\
\hline
$\Gamma[T_{cc}^+\to D^+ D^0 \gamma]$ & 6.1& 2.8& 1.9 \\
\hline
$\Gamma[T_{cc}^+]$ & 52 & 38 & 58\\ 
\hline
\end{tabular}
\caption{Partial and total widths in units of keV for three choices of $\theta$. The angle for $\Gamma_{\rm max}$ is chosen to maximize the total decay width for these channels.}
\label{table:widths}
\end{center}
\end{table}

\begin{figure}[t]
\centering
\includegraphics[trim={6.3cm 3.8cm 6.3cm 3.8cm},clip,scale=0.5]{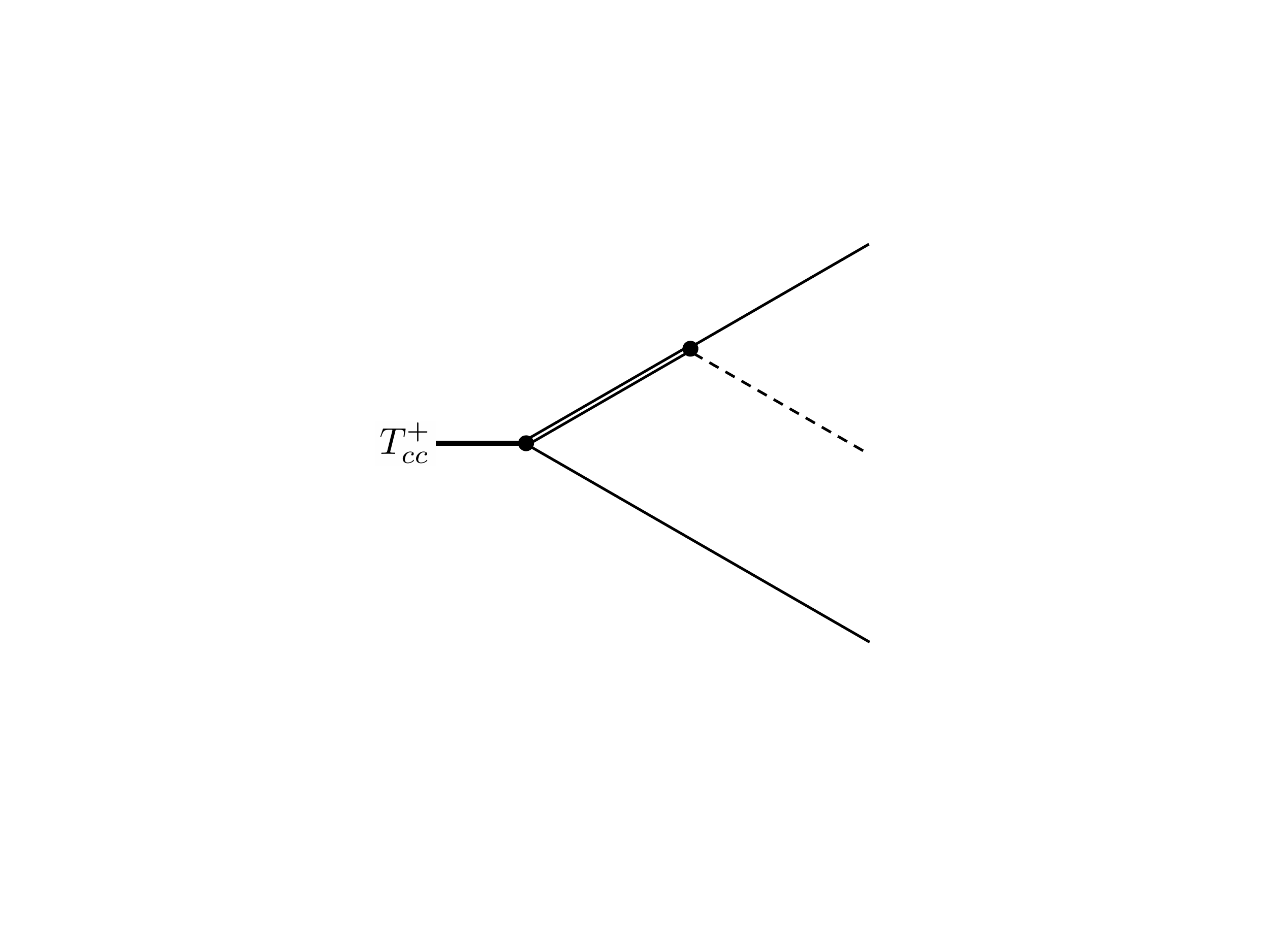}
\includegraphics[trim={6.3cm 3.8cm 6.3cm 3.8cm},clip,scale=0.5]{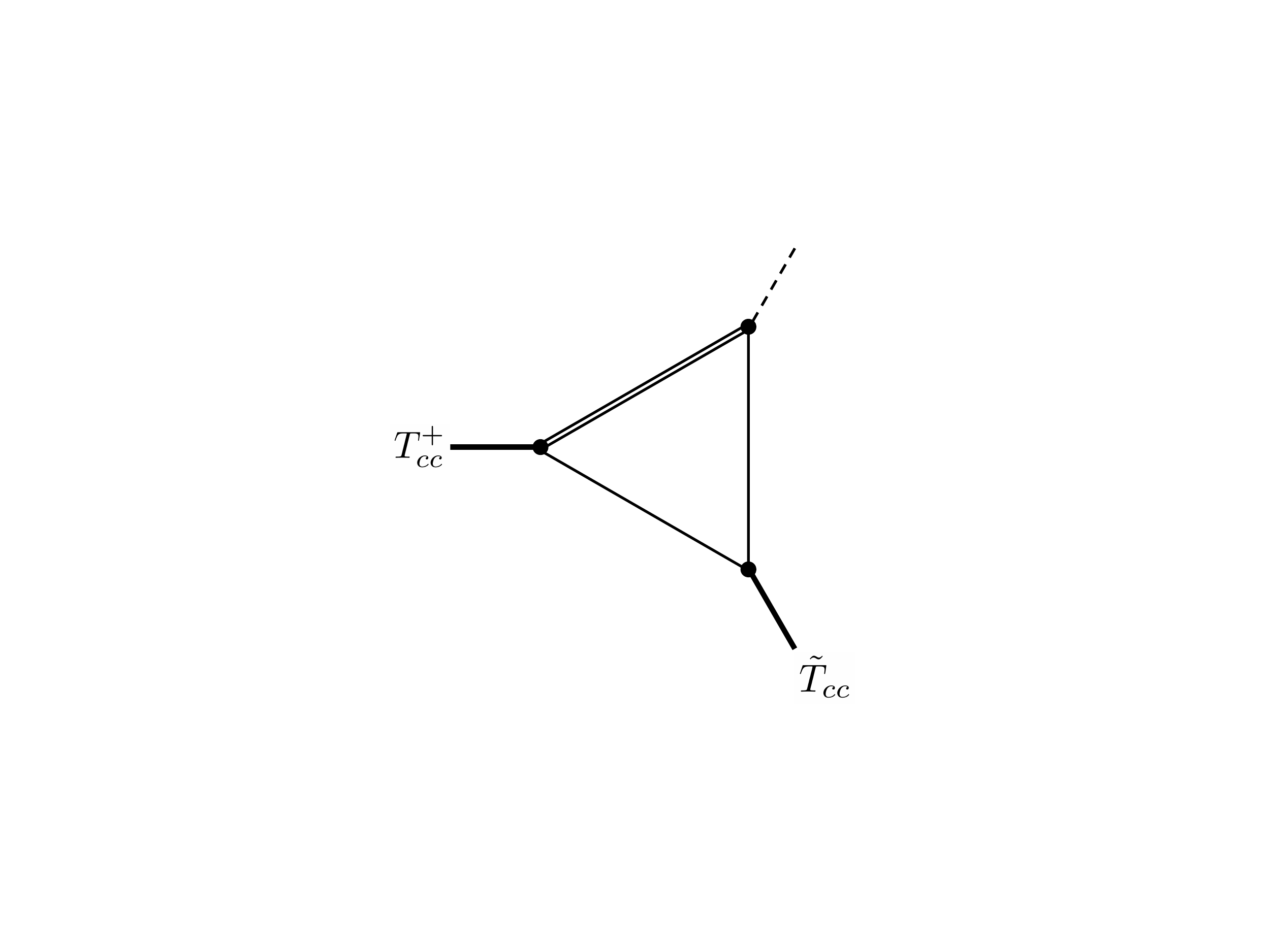}
\caption{The tree-level and one-loop  diagrams for the decay of the $T_{cc}^+$.  The thick lines represent $T_{cc}$ and $\tilde T_{cc}$ tetraquarks, the single thin lines represent $D$ mesons, the double lines represent $D^*$ mesons, and the dashed lines represent pions or photons.}
\label{feynmandiagrams}
\end{figure}

With all relevant terms in the Lagrangian described, obtaining the decay rates is straightforward.  The tree-level diagrams for the decays involve the $T_{cc}^+$ coupling to one of the two channels followed by a $D^*$ decay to a $D$ and a pion or a photon (see Fig. \ref{feynmandiagrams}).  For the strong and electromagnetic decay rates we find:
\begin{widetext}
\begin{subequations}
\bea
\frac{d\Gamma[T_{cc}^+\to D^0 D^0 \pi^+]}{dp_{D^0_1}^2 dp_{D^0_2}^2 } &=& c^2 _\theta\frac{g^2}{(4\pi f_\pi)^2}\frac{2 \gamma_0 p_\pi^2}{3}\left[\frac{1}{p_{D^0_1}^2+\gamma_0^2} +
\frac{1}{p_{D^0_2}^2+\gamma_0^2}\right]^2 \, ,  \label{DDpip}\\
\frac{d\Gamma[T_{cc}^+\to D^+ D^0 \pi^0]}{dp_{D^+}^2 dp_{D^0}^2 } &=& \frac{g^2}{(4\pi f_\pi)^2}\frac{2 p_\pi^2}{3 }\left[\frac{\sqrt{\gamma_0} \, c_\theta}{p_{D^+}^2+\gamma_0^2} -
\frac{\sqrt{\gamma_+} \,s_\theta}{p_{D^0}^2+\gamma_+^2}\right]^2\, , \label{DDpi0}
\\
\frac{d\Gamma[T_{cc}^+\to D^+ D^0 \gamma]}{dp_{D^+}^2 dp_{D^0}^2 } &=&  \frac{E_\gamma^2}{6 \pi^2 }\left[\frac{\sqrt{\gamma_0} \, c_\theta \mu_{D^0}}{p_{D^+}^2+\gamma_0^2} -
\frac{\sqrt{\gamma_+} \,s_\theta \mu_{D^+}}{p_{D^0}^2+\gamma_+^2}\right]^2 \, . \label{DDgamma}
\eea
\end{subequations}
\end{widetext}
In Eqs.~(\ref{DDpip}) and (\ref{DDpi0}), if we take $c_\theta = -s_\theta= 1/\sqrt{2}$ and $\gamma_0 = \gamma_+ = \gamma$ we recover the expression for the differential rate for $\chi_{c1}(3872) \to D^0 \bar{D}^0 \pi^0$
found in Ref.~\cite{Fleming:2007rp}. 
Integrating these expressions over three-body phase space yields the results 
shown in Table~\ref{table:widths}. The results are given for $\theta = -32.4^\circ$ ($I=0$), $\theta = +32.4^\circ$ ($I=1$), and $\theta=-8.34^\circ$, which turns out to maximize the total width of the three-body decays. We find the total strong decay widths are 47 keV, 36 keV, and 57 keV, respectively, and the electromagnetic decay widths are 6.1 keV, 2.8 keV, and 1.9 keV.  For the $I=0$ case, the total decay width for all three tree-level decays is about 52 keV, which is close to the width in the LHCb unitarized Breit-Wigner fit in Eq.~(\ref{BW_width2}).

Next we compare our results to previous theoretical calculations. Of the four papers previously cited for calculations of the decay~\cite{Meng:2021jnw,Ling:2021bir,Feijoo:2021ppq,Yan:2021wdl}, Ref.~\cite{Meng:2021jnw} uses methods that are most similar to ours. They solve a nonrelativistic coupled channel problem to infer the coupling of the $T_{cc}^+$ to the $DD^*$ states and evaluate diagrams that are identical to ours, albeit with relativistically invariant interactions. Since the charm mesons in the final state  are highly nonrelativistic, this should not be an important difference. They calculate the strong and radiative decay widths for all values of $\theta$ and their results  are consistent with ours. Reference~\cite{Ling:2021bir} convolves the same Feynman diagrams with a bound state wavefunction for the $T_{cc}^+$. They also fix the coupling of the $T_{cc}^+$ using the compositeness condition. Their predictions for the strong decay width and electromagnetic widths are very close to ours for the $I=0$ case. Reference~\cite{Feijoo:2021ppq} solves the Bethe-Salpeter equation for a potential between $D^*D$ mesons obtained from vector meson exchange and short distance interactions. These are tuned to produce a pole in the T-matrix at the mass of the $T_{cc}^+$. Feynman diagrams identical to ours with a relativistic Breit-Wigner for each $D^*$-meson propagator are evaluated and a strong decay width of 80 KeV, slightly larger than ours,  is extracted from the line shape. Reference~\cite{Yan:2021wdl}
convolves the $D^*D\pi$ amplitude with an effective range theory wavefunction to   calculate a strong decay width of about 50 keV at LO, which increases to 57 keV when next-to-leading-order (NLO) diagrams involving two-body operators are included. They also argue final state $D^0 D^0$ interactions can further increase the width.


\begin{figure*}[t]
\centering
\includegraphics[scale=1]{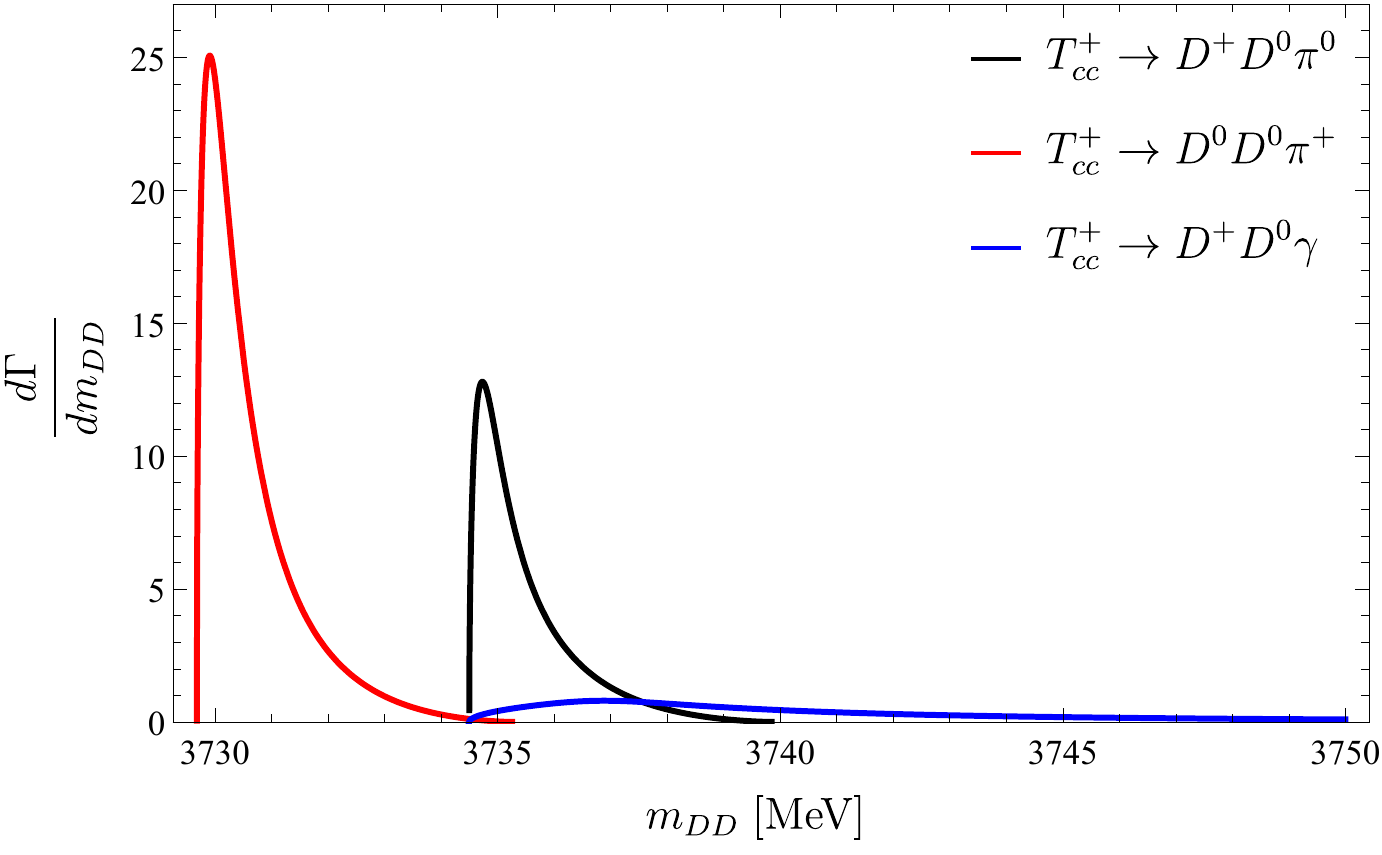}
\caption{Differential decay rates for $T_{cc}^+$ three-body decays.}
\label{invariantmass}
\end{figure*}

\begin{figure*}[t]
\centering
\includegraphics[scale=1]{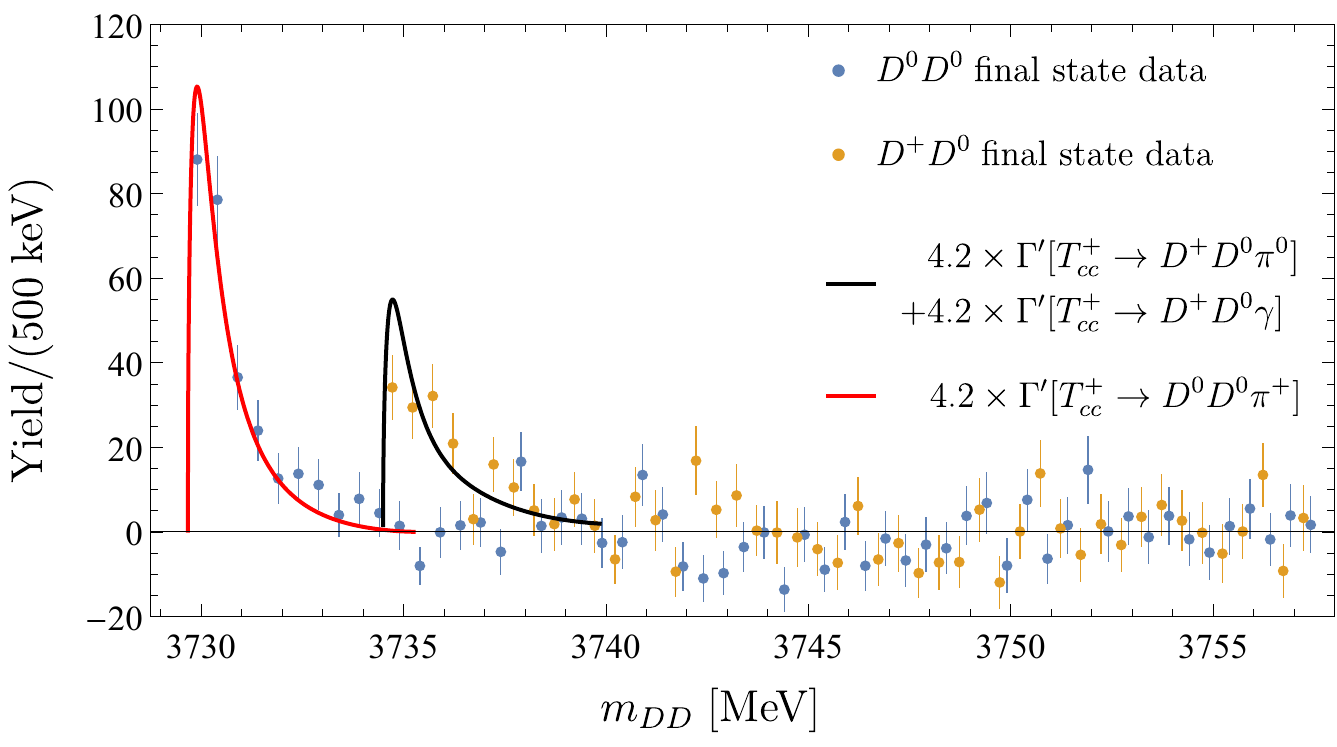}
\caption{The differential curves for $T_{cc}^+$ decays predicted by EFT superimposed on the binned experimental data from LHCb.  }
\label{invariantmass2}
\end{figure*}

We now turn to a brief discussion of uncertainties which will be qualitative. Reference~\cite{Dai:2019hrf} studied the decay $\Gamma[\chi_{c1}(3872)\to D^0 \bar{D}^0 \pi^0]$ including NLO corrections. There are several effects that appear in the NLO calculation: loops with pions, range corrections, and final state $\pi^0 D^0, \pi^0 \bar{D}^0$, and $D^0\bar{D}^0$ rescattering. Pion loops have a negligible effect on the width~\cite{Fleming:2007rp}. Information on $\pi D$ scattering lengths from the lattice~\cite{Liu:2012zya,Mohler_2013} is used to constrain $\pi^0 D^0$ and $\pi^0 \bar{D}^0$ rescattering. $D^0\bar{D}^0$ rescattering is unconstrained by experimental data or lattice simulations. The coefficient of a contact interaction mediating   
$D^0\bar{D}^0$ scattering is arbitrarily varied between $\pm 1\,\rm{fm}^2$ and this turns out to dominate the uncertainty. The total uncertainty in the decay rate is ${}^{+50\%}_{-30\%}$  when the binding energy of the $\chi_{c1}(3872)$ is 0.2 MeV.  We expect similar uncertainties in the prediction for the width of $T_{cc}^+$ as $D^0 D^0$ scattering is also not constrained. Also it is not clear that the range over which the contact interaction in Ref.~\cite{Dai:2019hrf} is varied is sufficiently large. The quoted range assumes the $D^0\bar{D}^0$ scattering length is not  larger than 1 fm, and hadron scattering lengths have been observed to be larger than that in many cases. Finally, extrapolating from the calculations in Ref.~\cite{Dai:2019hrf} is not totally straightforward because there is no coupled channel problem in that case. It would be interesting to perform a study similar to Ref.~\cite{Dai:2019hrf} for $T_{cc}^+$ and also obtain constraints on charm meson scattering lengths from the lattice. 

Reference~\cite{Dai:2019hrf} argues that while the partial width suffers from considerable uncertainty, the prediction of the energy spectrum of the pion in the final state is robust, up to normalization. The spectrum is highly peaked near maximal energy and the location of the peak and shape of the distribution are not affected by NLO effects. They also observed that the location of the peak in the distribution is sensitive to the binding of the $\chi_{c1}(3872)$, which provides an alternative way of extracting this quantity from data.
In this paper, for the case $I=0$, we plot the differential decay rates for the three decays as a function of the invariant mass of the $DD$ pair, $m_{DD}$, in Fig.~\ref{invariantmass}.  Note that maximal pion or photon energy corresponds to minimal $m_{DD}$. 
By analogy with similar conclusions about the $\chi_{c1}(3872)$ in XEFT~\cite{Dai:2019hrf}, the sharpness of the peaks is due to the molecular nature of the $T_{cc}^+$. The decay is via an intermediate state with a virtual $D^*$ and the pole in the propagator is responsible for the sharpness of those peaks in the strong decays.  The peak is not so pronounced in the electromagnetic decay.

In Fig.~\ref{invariantmass2} we compare our results to data on the invariant mass spectra from Fig.~4 of Ref.~\cite{LHCb:2021auc}. We have taken the experimental data and subtracted the background extracted from that figure and superimposed our $m_{D^0D^0}$ and $m_{D^+D^0}$ distributions. For the $m_{D^+D^0}$ distribution we combine the contributions from $T_{cc}^+\to D^+D^0\pi^0$
and  $T_{cc}^+\to D^+D^0\gamma$.   The curves are each scaled by a factor of $4.2$ so that the peak of our $m_{D^0D^0}$ distribution is approximately the same height
as the peak of the $m_{D^0D^0}$ distribution extracted by LHCb~\cite{LHCb:2021auc}. 
We see that the shapes and relative normalization are in very good agreement with the mass distributions observed by LHCb, especially for $D^0D^0$. 


\begin{figure*}[t]
\centering
\begin{minipage}{0.33\textwidth}
\centering
\includegraphics[scale=0.6]{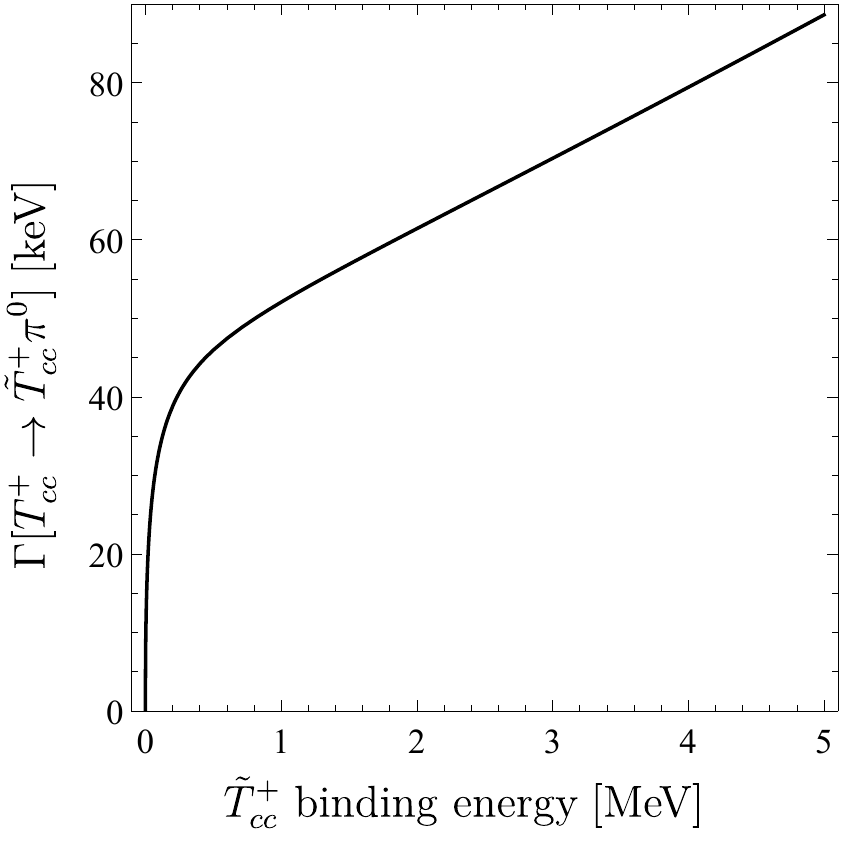}
\label{LoopPi01}
\end{minipage}%
\begin{minipage}{0.33\textwidth}
\centering
\includegraphics[scale=0.6]{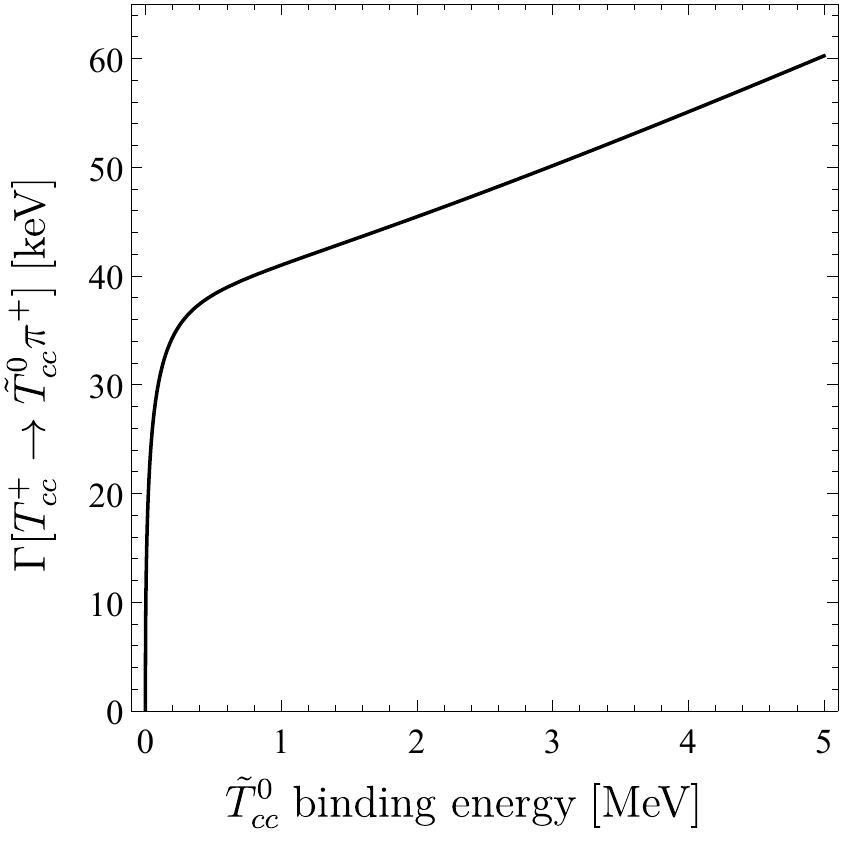}
\label{LoopPiPlus1}
\end{minipage}%
\begin{minipage}{0.33\textwidth}
\centering
\includegraphics[scale=0.6]{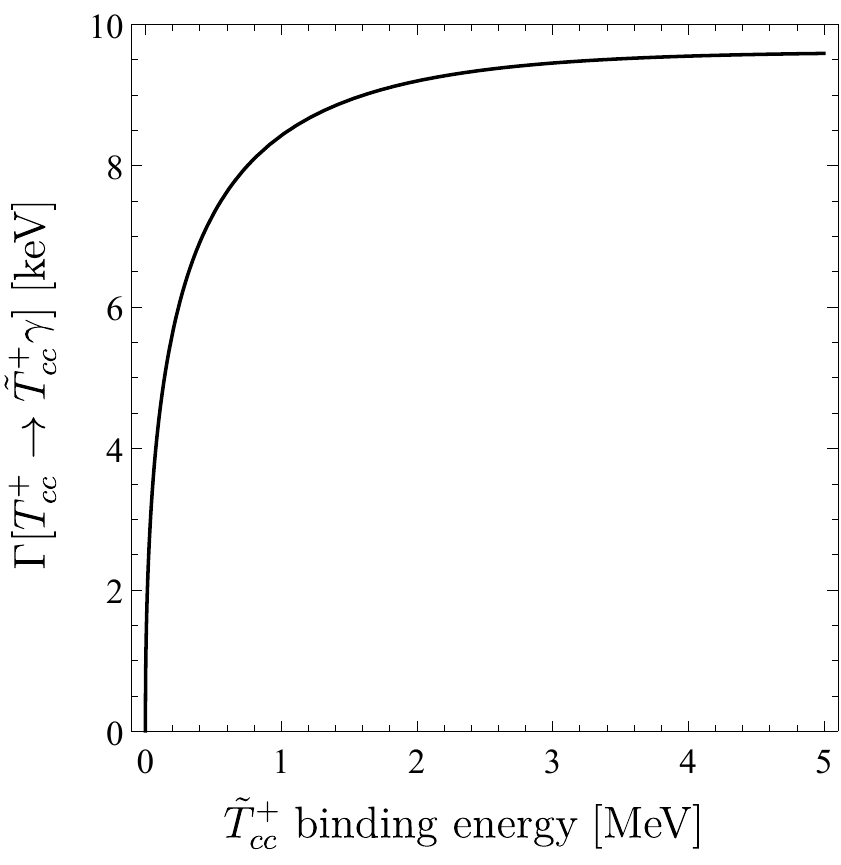}
\label{LoopPhoton1}
\end{minipage}
\caption{Decay rates for the $T_{cc}^+$ to $\tilde{T}_{cc}$ and a pion/photon as a function of the $\tilde{T}_{cc}$ binding energy for $\theta=-32.4^\circ$.}
\label{LoopPlots1}
\end{figure*}

Finally we consider the possibility that in addition to $T_{cc}^+$ there are also shallow bound states of 
$D^0D^+\;(\tilde{T}_{cc}^+)$ and 
$D^0 D^0\;(\tilde{T}_{cc}^0)$.  As stated earlier, the total decay rate in the $I=0$ channel for the three three-body processes is  $52 \; \text{keV}$, which agrees well with the LHCb width in Eq.~(\ref{BW_width2}), but is considerably lower than the width in Eq.~(\ref{BW_width}).  The large discrepancy between the two fits to the experimental data makes it worthwhile to consider other possible decay channels which could increase the predicted decay width of the $T_{cc}^+$. So we will consider two-body decays involving the $\tilde{T}_{cc}$ bound states.
Reviewing the theory predictions collected in Ref.~\cite{LHCb:2021vvq} most $I=1$ $J^P=0^+$ doubly charm tetraquarks are predicted to be a few hundred MeV above the $D D$ threshold,   though Refs.~\cite{Tan_2020,PhysRevD.80.114023} have predictions for an $I=1$ $J^P=0^+$ tetraquark less than 1.5 MeV above threshold.  Given theoretical uncertainties it seems at least conceivable that shallow bound states of pseudoscalar charm mesons  could exist.    The triangle diagram mediating the decays $T_{cc}^+\to \tilde{T}_{cc} \pi, \tilde{T}_{cc} \gamma$ is shown in Fig.~\ref{feynmandiagrams}. Using the same coupled-channel analysis used for the three-body decays, the  rates for two-body strong and electromagnetic decays are
\begin{widetext}
\begin{subequations}
\bea
\Gamma[T_{cc}^+ \rightarrow \tilde{T}_{cc}^+ \pi^0] &=& 
\frac{| {\bf p}_\pi|m_{\tilde{T}}}{6 \pi m_T}
\bigg(\frac{g}{\sqrt{2}f_\pi}\bigg)^2 
\big|c_\theta F(m_{D^0},m_{D^{*+}},m_{D^+},m_{\tilde{T}}) - s_\theta F(m_{D^+},m_{D^{*0}},m_{D^0},m_{\tilde{T}})\big|^2 \, ,  \\
\Gamma[T_{cc}^+ \rightarrow \tilde{T}_{cc}^0 \pi^+] &=& \frac{| {\bf p}_\pi|m_{\tilde{T}}}{6 \pi m_T}\bigg(\frac{g}{f_\pi}\bigg)^2\big|c_\theta F(m_{D^0},m_{D^{*+}},m_{D^0},m_{\tilde{T}})\big|^2 \, ,  \\
\Gamma[T_{cc}^+ \rightarrow \tilde{T}_{cc}^+ \gamma] &=& 
\frac{| {\bf p}_\gamma|m_{\tilde{T}}}{3 \pi m_T}
\big|\mu_{D^+}c_\theta F(m_{D^0},m_{D^{*+}},m_{D^+},m_{\tilde{T}}) + \mu_{D^0}s_\theta F(m_{D^+},m_{D^{*0}},m_{D^0},m_{\tilde{T}})\big|^2 \, .
\eea
\end{subequations}
\end{widetext}
Here $m_{\tilde T}$ is the mass of the $\tilde T_{cc}$ in the decay and 
the function $F(m_1,m_2,m_3,m_{\tilde{T}})$ comes from evaluating the triangle diagram.  The function $F(m_1,m_2,m_3,m_{\tilde{T}})$ and its various parameters are given by

\begin{widetext}
\bea
F(m_1,m_2,m_3,m_{\tilde{T}}) &=& \sqrt{\frac{\gamma_{12}\gamma_{13}}{b^2}}\bigg[ \tan^{-1}\bigg(\frac{c_2-c_1}{2\sqrt{c_2b^2{\bf p}_{\pi}^2}}\bigg) + \tan^{-1} \bigg(\frac{2b^2{\bf p}_\pi^2+c_1-c_2}{2\sqrt{b^2{\bf p}_\pi^2(c_2-b^2{\bf p}_\pi^2)}}\bigg)\bigg], \nn \\
\gamma_{12} &=&\sqrt{-2\mu_{12}(m_T-m_1-m_2)} \qquad
\gamma_{13} =\sqrt{-2\mu_{13}(E_{\tilde{T}}-m_1-m_3)} \nonumber \\
\mu_{ij}^{-1} &=& m_i^{-1}+m_j^{-1} \qquad
c_1 = \gamma^2_{12}\qquad
c_2 = \frac{\mu_{13}}{m_3}{\bf p}_\pi^2+ \gamma_{13}^2 \qquad
b =\mu_{13}/m_3. \qquad
\eea
\end{widetext}

\begin{figure*}[t]
\centering
\begin{minipage}{0.33\textwidth}
\centering
\includegraphics[scale=0.6]{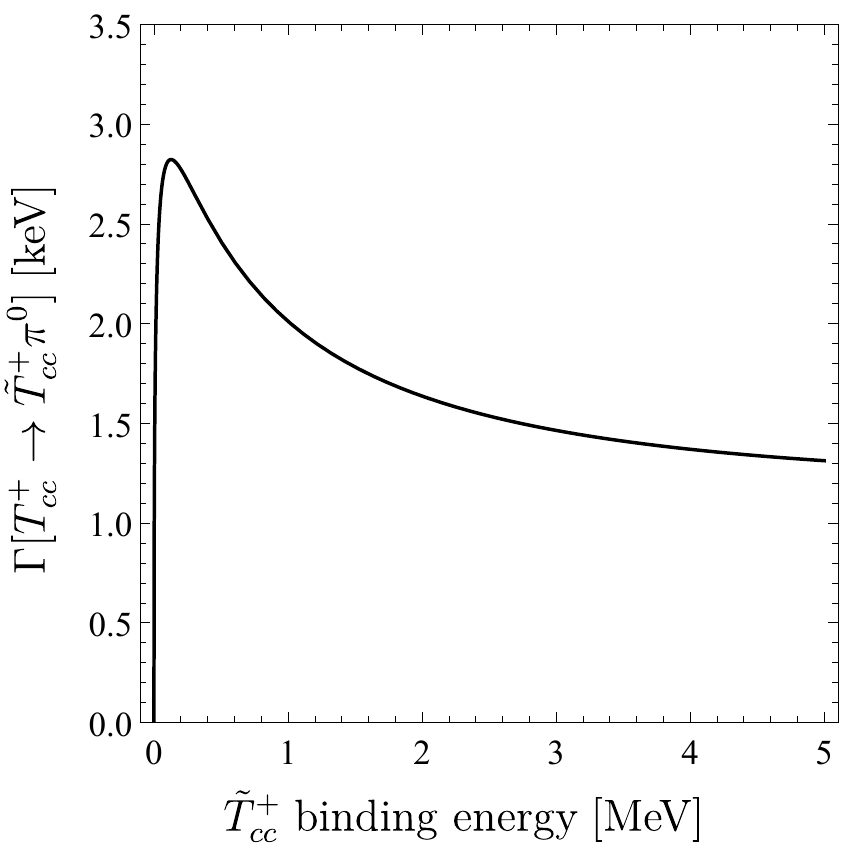}
\label{LoopPi02}
\end{minipage}%
\begin{minipage}{0.33\textwidth}
\centering
\includegraphics[scale=0.6]{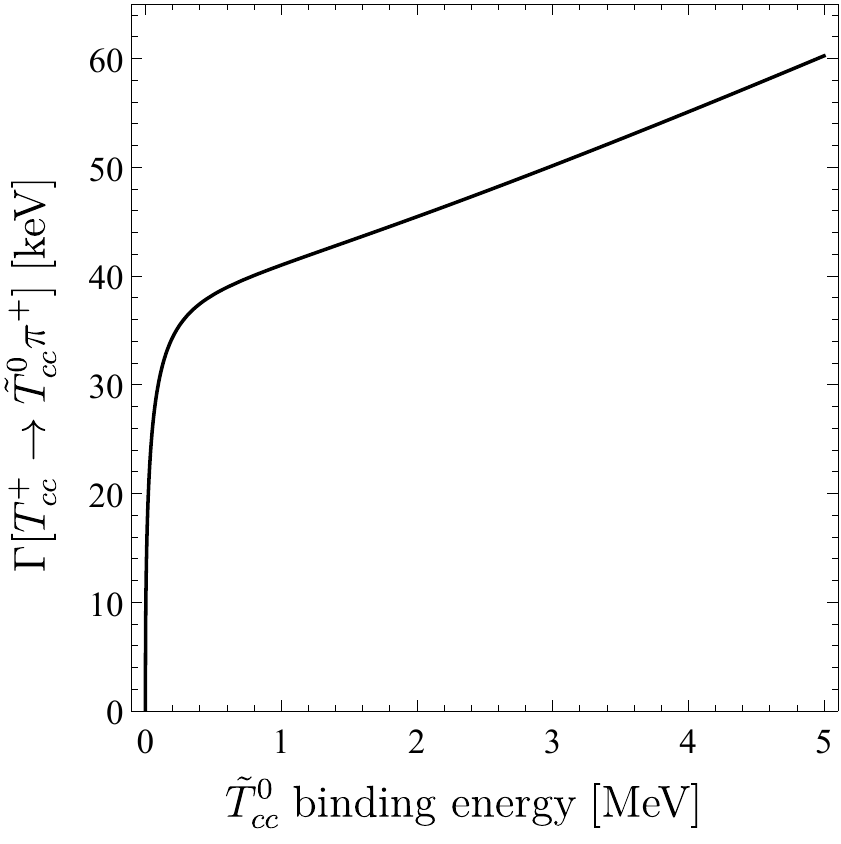}
\label{LoopPiPlus2}
\end{minipage}%
\begin{minipage}{0.33\textwidth}
\centering
\includegraphics[scale=0.6]{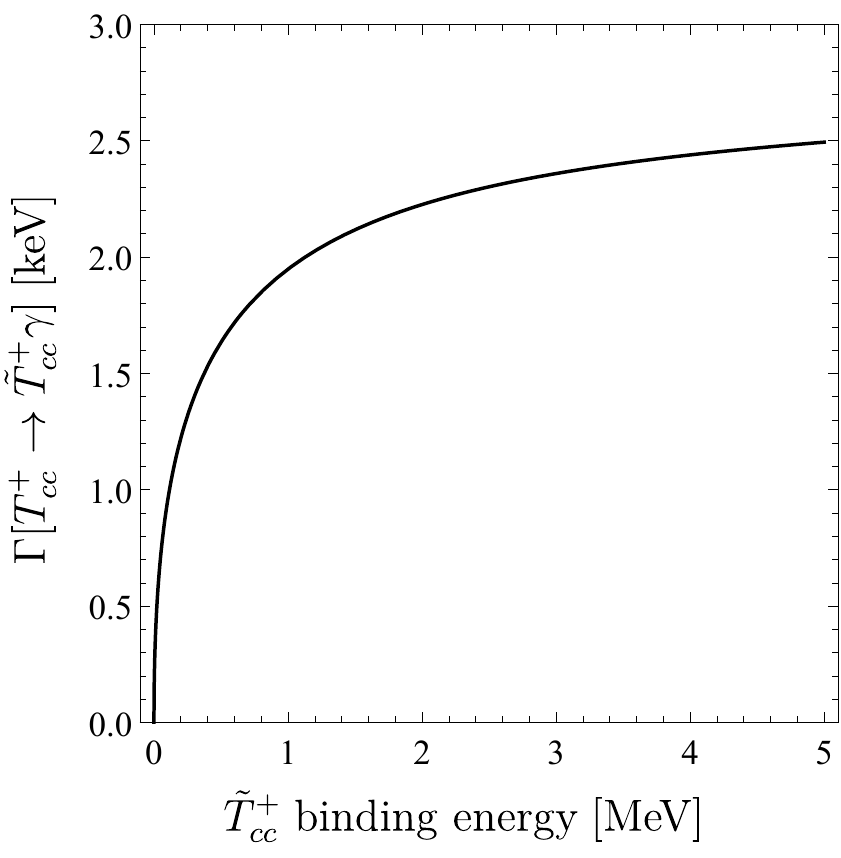}
\label{LoopPhoton2}
\end{minipage}
\caption{Decay rates for the $T_{cc}^+$ to $\tilde{T}_{cc}$ and a pion/photon as a function of the $\tilde{T}_{cc}$ binding energy for $\theta=+32.4^\circ$.}
\label{LoopPlots2}
\end{figure*}

\begin{figure*}[t]
\centering
\begin{minipage}{0.33\textwidth}
\centering
\includegraphics[scale=0.6]{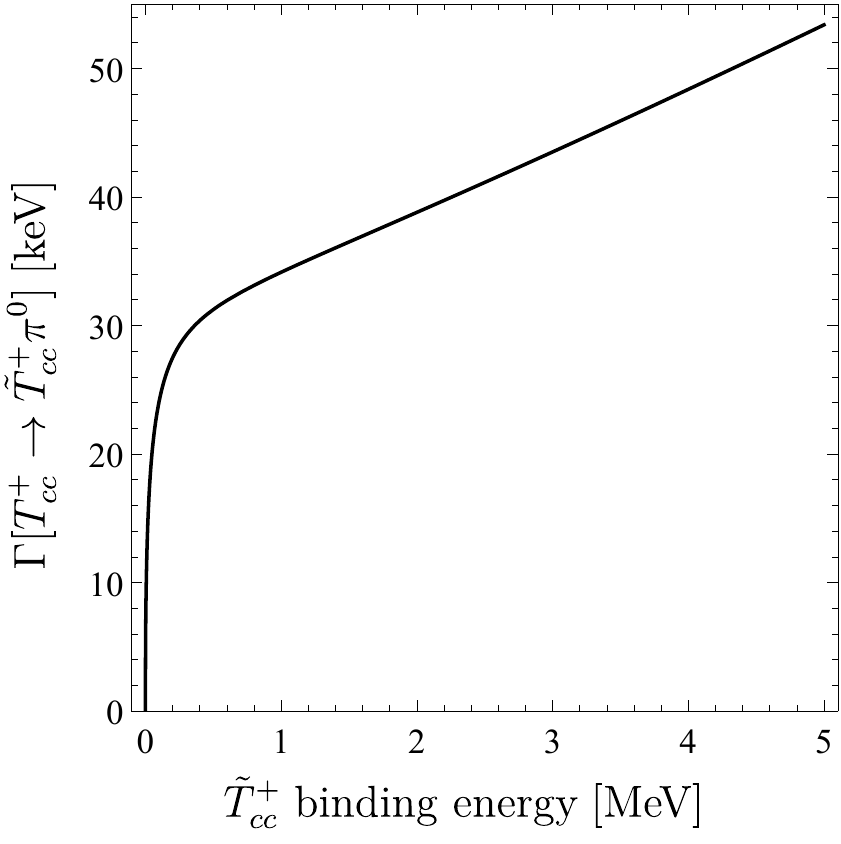}
\label{LoopPi03}
\end{minipage}%
\begin{minipage}{0.33\textwidth}
\centering
\includegraphics[scale=0.6]{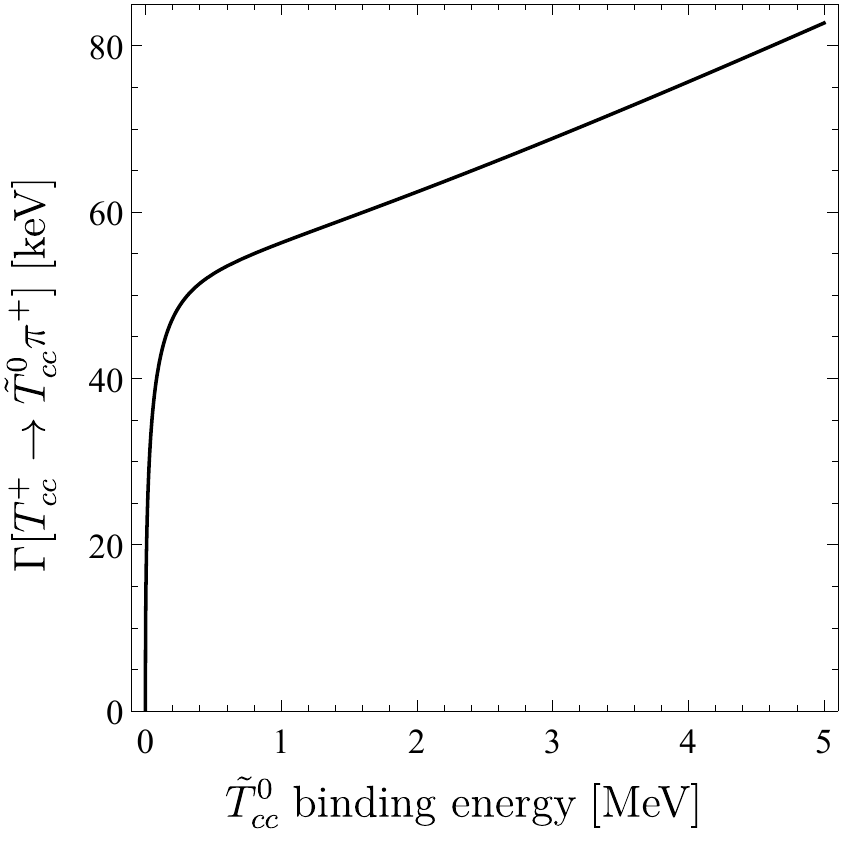}
\label{LoopPiPlus3}
\end{minipage}%
\begin{minipage}{0.33\textwidth}
\centering
\includegraphics[scale=0.6]{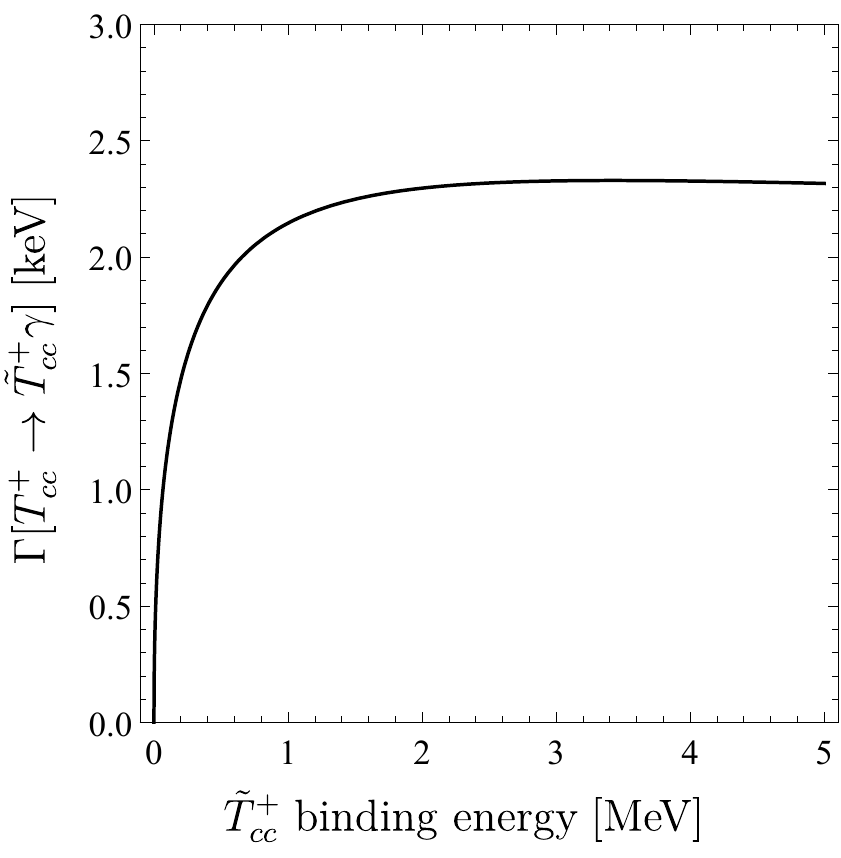}
\label{LoopPhoton3}
\end{minipage}
\caption{Decay rates for the $T_{cc}^+$ to $\tilde{T}_{cc}$ and a pion/photon as a function of the $\tilde{T}_{cc}$ binding energy for $\theta=-8.34^\circ$.}
\label{LoopPlots3}
\end{figure*}

We plot the decay rate for these processes as a function of the binding energy of the ${\tilde{T}}_{cc}^{+/0}$.  In the $I=0$ channel, shown in Fig. \ref{LoopPlots1}, the decay rate strictly increases with the binding energy in the domain $[0,5] \; \text{MeV}$.  The $I=1$ channel is shown in  Fig. \ref{LoopPlots2}; we see that the neutral pion decay is greatly suppressed and has a local maximum at very small binding energy. This is presumably because of an accidental cancellation between the two amplitudes for this process at this particular angle. If we choose $\theta=-8.34^\circ$, shown in Fig.~\ref{LoopPlots3}, the plots of the width as a function of binding energy are very similar to Fig.~\ref{LoopPlots1}. In the $I=0$ channel the decay width of the $T_{cc}^+$ could be enhanced by as much as $150 \; \text{keV}$ if the binding energy of  ${\tilde{T}}_{cc}^{0}$ and ${\tilde{T}}_{cc}^{+}$ are as large as 5 MeV. This would bring the EFT prediction for the width much closer to the experimental result in Eq.~(\ref{BW_width}), including the uncertainty, albeit far above the unitarized Breit-Wigner result in Eq. (\ref{BW_width2}). If the binding energy of these states is the same as $T_{cc}^+$, the width will be increased by about 80 keV.  Note the effect of  two-body decays to bound states on the total decay rate is much larger than the error estimate of $D^0 \bar{D^0}$ rescattering effects in Ref.~\cite{Dai:2019hrf}.
The range of values for the contact interaction mediating $D^0\bar{D}^0$ scattering in Ref.~\cite{Dai:2019hrf} does not include scattering lengths that are sufficiently large to accommodate shallow bound states.

To summarize, we have developed an EFT for the $T_{cc}^+$ which is applicable if it is primarily a molecular state coupled to the $D^0D^{*+}$ and $D^+D^{*0}$ channels. This theory is used to calculate the partial widths for the decays $T_{cc}^+ \to D^0 D^0\pi^+, D^+ D^0\pi^0$ and $D^+D^0\gamma$. The calculated total width from these three decays is close to the value of the width extracted from fitting experimental data with a unitarized Breit-Wigner line shape~\cite{LHCb:2021auc}. Furthermore, our calculations are consistent with other theoretical calculations assuming a molecular interpretation of $T_{cc}^+$~\cite{Meng:2021jnw,Ling:2021bir,Feijoo:2021ppq,Yan:2021wdl}. We emphasize that the uncertainties on the total width due to final state rescattering effects are potentially quite large by considering a NLO  calculation for a similar $\chi_{c1}(3872)$ decay in Ref.~\cite{Dai:2019hrf}. Our EFT predictions for the differential spectra as a function of the invariant mass of the $DD$ pair show good agreement with the LHCb mass distributions~\cite{LHCb:2021auc}. Finally, we entertained the possibility of shallow bound states of pseudoscalar charm mesons. If such states exist, the two-body decay of $T_{cc}^+$ to these bound states would greatly enhance its width. If the width of $T_{cc}^+$ is confirmed to be $\approx 50$ keV, our calculation provides strong evidence that these states do not exist. 


{\bf Acknowledgments} - We thank M. Mikhasenko for correspondence about the experimental results.  T. M. and R. H. thank I.~Low and A.~Mohapatra for helpful discussions, and L. Dai for supplying a \emph{Mathematica} code which was modified to produce the results in this paper. The authors also thank M. P.~Valderrama for a careful reading of the manuscript as well as insightful discussions.
T.~M. and R.~H. are supported by the U.S. Department of Energy, Office of Science, Office of Nuclear Physics under grant Contract Numbers  DE-FG02-05ER41367. S.~F. is supported by the U.S. Department of Energy, Office of Science, Office of Nuclear Physics, under award number DE-FG02-04ER41338.

\bibliography{main}

\providecommand{\noopsort}[1]{}\providecommand{\singleletter}[1]{#1}%
\begin{thebibliography}{73}%
\makeatletter
\providecommand \@ifxundefined [1]{%
 \@ifx{#1\undefined}
}%
\providecommand \@ifnum [1]{%
 \ifnum #1\expandafter \@firstoftwo
 \else \expandafter \@secondoftwo
 \fi
}%
\providecommand \@ifx [1]{%
 \ifx #1\expandafter \@firstoftwo
 \else \expandafter \@secondoftwo
 \fi
}%
\providecommand \natexlab [1]{#1}%
\providecommand \enquote  [1]{``#1''}%
\providecommand \bibnamefont  [1]{#1}%
\providecommand \bibfnamefont [1]{#1}%
\providecommand \citenamefont [1]{#1}%
\providecommand \href@noop [0]{\@secondoftwo}%
\providecommand \href [0]{\begingroup \@sanitize@url \@href}%
\providecommand \@href[1]{\@@startlink{#1}\@@href}%
\providecommand \@@href[1]{\endgroup#1\@@endlink}%
\providecommand \@sanitize@url [0]{\catcode `\\12\catcode `\$12\catcode
  `\&12\catcode `\#12\catcode `\^12\catcode `\_12\catcode `\%12\relax}%
\providecommand \@@startlink[1]{}%
\providecommand \@@endlink[0]{}%
\providecommand \url  [0]{\begingroup\@sanitize@url \@url }%
\providecommand \@url [1]{\endgroup\@href {#1}{\urlprefix }}%
\providecommand \urlprefix  [0]{URL }%
\providecommand \Eprint [0]{\href }%
\providecommand \doibase [0]{http://dx.doi.org/}%
\providecommand \selectlanguage [0]{\@gobble}%
\providecommand \bibinfo  [0]{\@secondoftwo}%
\providecommand \bibfield  [0]{\@secondoftwo}%
\providecommand \translation [1]{[#1]}%
\providecommand \BibitemOpen [0]{}%
\providecommand \bibitemStop [0]{}%
\providecommand \bibitemNoStop [0]{.\EOS\space}%
\providecommand \EOS [0]{\spacefactor3000\relax}%
\providecommand \BibitemShut  [1]{\csname bibitem#1\endcsname}%
\let\auto@bib@innerbib\@empty
\bibitem [{\citenamefont {Aaij}\ \emph
  {et~al.}(2021{\natexlab{a}})\citenamefont {Aaij} \emph
  {et~al.}}]{LHCb:2021vvq}%
  \BibitemOpen
  \bibfield  {author} {\bibinfo {author} {\bibfnamefont {R.}~\bibnamefont
  {Aaij}} \emph {et~al.} (\bibinfo {collaboration} {LHCb}),\ }\href@noop {} {\
  (\bibinfo {year} {2021}{\natexlab{a}})},\ \Eprint
  {http://arxiv.org/abs/2109.01038} {arXiv:2109.01038 [hep-ex]} \BibitemShut
  {NoStop}%
\bibitem [{\citenamefont {Aaij}\ \emph
  {et~al.}(2021{\natexlab{b}})\citenamefont {Aaij} \emph
  {et~al.}}]{LHCb:2021auc}%
  \BibitemOpen
  \bibfield  {author} {\bibinfo {author} {\bibfnamefont {R.}~\bibnamefont
  {Aaij}} \emph {et~al.} (\bibinfo {collaboration} {LHCb}),\ }\href@noop {} {\
  (\bibinfo {year} {2021}{\natexlab{b}})},\ \Eprint
  {http://arxiv.org/abs/2109.01056} {arXiv:2109.01056 [hep-ex]} \BibitemShut
  {NoStop}%
\bibitem [{\citenamefont {Muheim}(2021)}]{Muheim}%
  \BibitemOpen
  \bibfield  {author} {\bibinfo {author} {\bibfnamefont {F.}~\bibnamefont
  {Muheim}},\ }\href {https://indico.desy.de/event/28202/contributions/102717/}
  {\  (\bibinfo {year} {2021})},\ \bibinfo {note} {the European Physical
  Society Conference on High Energy Physics}\BibitemShut {NoStop}%
\bibitem [{\citenamefont {Polyakov}(2021)}]{Polyakov}%
  \BibitemOpen
  \bibfield  {author} {\bibinfo {author} {\bibfnamefont {I.}~\bibnamefont
  {Polyakov}},\ }\href
  {https://indico.desy.de/event/28202/contributions/105627/} {\  (\bibinfo
  {year} {2021})},\ \bibinfo {note} {the European Physical Society Conference
  on High Energy Physics}\BibitemShut {NoStop}%
\bibitem [{\citenamefont {An}(2021)}]{An}%
  \BibitemOpen
  \bibfield  {author} {\bibinfo {author} {\bibfnamefont {L.}~\bibnamefont
  {An}},\ }\href
  {https://indico.nucleares.unam.mx/event/1541/session/4/contribution/35/material/slides/0.pdf}
  {\  (\bibinfo {year} {2021})},\ \bibinfo {note} {19th International
  Conference on Hadron Spectroscopy and Structure}\BibitemShut {NoStop}%
\bibitem [{\citenamefont {Dias}\ \emph {et~al.}(2011)\citenamefont {Dias},
  \citenamefont {Narison}, \citenamefont {Navarra}, \citenamefont {Nielsen},\
  and\ \citenamefont {Richard}}]{Dias:2011mi}%
  \BibitemOpen
  \bibfield  {author} {\bibinfo {author} {\bibfnamefont {J.~M.}\ \bibnamefont
  {Dias}}, \bibinfo {author} {\bibfnamefont {S.}~\bibnamefont {Narison}},
  \bibinfo {author} {\bibfnamefont {F.~S.}\ \bibnamefont {Navarra}}, \bibinfo
  {author} {\bibfnamefont {M.}~\bibnamefont {Nielsen}}, \ and\ \bibinfo
  {author} {\bibfnamefont {J.~M.}\ \bibnamefont {Richard}},\ }\href {\doibase
  10.1016/j.physletb.2011.07.082} {\bibfield  {journal} {\bibinfo  {journal}
  {Phys. Lett. B}\ }\textbf {\bibinfo {volume} {703}},\ \bibinfo {pages} {274}
  (\bibinfo {year} {2011})},\ \Eprint {http://arxiv.org/abs/1105.5630}
  {arXiv:1105.5630 [hep-ph]} \BibitemShut {NoStop}%
\bibitem [{\citenamefont {Janc}\ and\ \citenamefont
  {Rosina}(2004)}]{Janc:2004qn}%
  \BibitemOpen
  \bibfield  {author} {\bibinfo {author} {\bibfnamefont {D.}~\bibnamefont
  {Janc}}\ and\ \bibinfo {author} {\bibfnamefont {M.}~\bibnamefont {Rosina}},\
  }\href {\doibase 10.1007/s00601-004-0068-9} {\bibfield  {journal} {\bibinfo
  {journal} {Few Body Syst.}\ }\textbf {\bibinfo {volume} {35}},\ \bibinfo
  {pages} {175} (\bibinfo {year} {2004})},\ \Eprint
  {http://arxiv.org/abs/hep-ph/0405208} {arXiv:hep-ph/0405208} \BibitemShut
  {NoStop}%
\bibitem [{\citenamefont {Zouzou}\ \emph {et~al.}(1986)\citenamefont {Zouzou},
  \citenamefont {Silvestre-Brac}, \citenamefont {Gignoux},\ and\ \citenamefont
  {Richard}}]{Zouzou:1986qh}%
  \BibitemOpen
  \bibfield  {author} {\bibinfo {author} {\bibfnamefont {S.}~\bibnamefont
  {Zouzou}}, \bibinfo {author} {\bibfnamefont {B.}~\bibnamefont
  {Silvestre-Brac}}, \bibinfo {author} {\bibfnamefont {C.}~\bibnamefont
  {Gignoux}}, \ and\ \bibinfo {author} {\bibfnamefont {J.~M.}\ \bibnamefont
  {Richard}},\ }\href {\doibase 10.1007/BF01557611} {\bibfield  {journal}
  {\bibinfo  {journal} {Z. Phys. C}\ }\textbf {\bibinfo {volume} {30}},\
  \bibinfo {pages} {457} (\bibinfo {year} {1986})}\BibitemShut {NoStop}%
\bibitem [{\citenamefont {Vijande}\ \emph {et~al.}(2006)\citenamefont
  {Vijande}, \citenamefont {Valcarce},\ and\ \citenamefont
  {Tsushima}}]{Vijande:2006jf}%
  \BibitemOpen
  \bibfield  {author} {\bibinfo {author} {\bibfnamefont {J.}~\bibnamefont
  {Vijande}}, \bibinfo {author} {\bibfnamefont {A.}~\bibnamefont {Valcarce}}, \
  and\ \bibinfo {author} {\bibfnamefont {K.}~\bibnamefont {Tsushima}},\ }\href
  {\doibase 10.1103/PhysRevD.74.054018} {\bibfield  {journal} {\bibinfo
  {journal} {Phys. Rev. D}\ }\textbf {\bibinfo {volume} {74}},\ \bibinfo
  {pages} {054018} (\bibinfo {year} {2006})},\ \Eprint
  {http://arxiv.org/abs/hep-ph/0608316} {arXiv:hep-ph/0608316} \BibitemShut
  {NoStop}%
\bibitem [{\citenamefont {Brink}\ and\ \citenamefont
  {Stancu}(1998)}]{Brink:1998as}%
  \BibitemOpen
  \bibfield  {author} {\bibinfo {author} {\bibfnamefont {D.~M.}\ \bibnamefont
  {Brink}}\ and\ \bibinfo {author} {\bibfnamefont {F.}~\bibnamefont {Stancu}},\
  }\href {\doibase 10.1103/PhysRevD.57.6778} {\bibfield  {journal} {\bibinfo
  {journal} {Phys. Rev. D}\ }\textbf {\bibinfo {volume} {57}},\ \bibinfo
  {pages} {6778} (\bibinfo {year} {1998})}\BibitemShut {NoStop}%
\bibitem [{\citenamefont {Swanson}(2006)}]{Swanson:2006st}%
  \BibitemOpen
  \bibfield  {author} {\bibinfo {author} {\bibfnamefont {E.~S.}\ \bibnamefont
  {Swanson}},\ }\href {\doibase 10.1016/j.physrep.2006.04.003} {\bibfield
  {journal} {\bibinfo  {journal} {Phys. Rept.}\ }\textbf {\bibinfo {volume}
  {429}},\ \bibinfo {pages} {243} (\bibinfo {year} {2006})},\ \Eprint
  {http://arxiv.org/abs/hep-ph/0601110} {arXiv:hep-ph/0601110} \BibitemShut
  {NoStop}%
\bibitem [{\citenamefont {Navarra}\ \emph {et~al.}(2007)\citenamefont
  {Navarra}, \citenamefont {Nielsen},\ and\ \citenamefont
  {Lee}}]{Navarra:2007yw}%
  \BibitemOpen
  \bibfield  {author} {\bibinfo {author} {\bibfnamefont {F.~S.}\ \bibnamefont
  {Navarra}}, \bibinfo {author} {\bibfnamefont {M.}~\bibnamefont {Nielsen}}, \
  and\ \bibinfo {author} {\bibfnamefont {S.~H.}\ \bibnamefont {Lee}},\ }\href
  {\doibase 10.1016/j.physletb.2007.04.010} {\bibfield  {journal} {\bibinfo
  {journal} {Phys. Lett. B}\ }\textbf {\bibinfo {volume} {649}},\ \bibinfo
  {pages} {166} (\bibinfo {year} {2007})},\ \Eprint
  {http://arxiv.org/abs/hep-ph/0703071} {arXiv:hep-ph/0703071} \BibitemShut
  {NoStop}%
\bibitem [{\citenamefont {Du}\ \emph {et~al.}(2013)\citenamefont {Du},
  \citenamefont {Chen}, \citenamefont {Chen},\ and\ \citenamefont
  {Zhu}}]{Du:2012wp}%
  \BibitemOpen
  \bibfield  {author} {\bibinfo {author} {\bibfnamefont {M.-L.}\ \bibnamefont
  {Du}}, \bibinfo {author} {\bibfnamefont {W.}~\bibnamefont {Chen}}, \bibinfo
  {author} {\bibfnamefont {X.-L.}\ \bibnamefont {Chen}}, \ and\ \bibinfo
  {author} {\bibfnamefont {S.-L.}\ \bibnamefont {Zhu}},\ }\href {\doibase
  10.1103/PhysRevD.87.014003} {\bibfield  {journal} {\bibinfo  {journal} {Phys.
  Rev. D}\ }\textbf {\bibinfo {volume} {87}},\ \bibinfo {pages} {014003}
  (\bibinfo {year} {2013})},\ \Eprint {http://arxiv.org/abs/1209.5134}
  {arXiv:1209.5134 [hep-ph]} \BibitemShut {NoStop}%
\bibitem [{\citenamefont {Ebert}\ \emph {et~al.}(2007)\citenamefont {Ebert},
  \citenamefont {Faustov}, \citenamefont {Galkin},\ and\ \citenamefont
  {Lucha}}]{Ebert:2007rn}%
  \BibitemOpen
  \bibfield  {author} {\bibinfo {author} {\bibfnamefont {D.}~\bibnamefont
  {Ebert}}, \bibinfo {author} {\bibfnamefont {R.~N.}\ \bibnamefont {Faustov}},
  \bibinfo {author} {\bibfnamefont {V.~O.}\ \bibnamefont {Galkin}}, \ and\
  \bibinfo {author} {\bibfnamefont {W.}~\bibnamefont {Lucha}},\ }\href
  {\doibase 10.1103/PhysRevD.76.114015} {\bibfield  {journal} {\bibinfo
  {journal} {Phys. Rev. D}\ }\textbf {\bibinfo {volume} {76}},\ \bibinfo
  {pages} {114015} (\bibinfo {year} {2007})},\ \Eprint
  {http://arxiv.org/abs/0706.3853} {arXiv:0706.3853 [hep-ph]} \BibitemShut
  {NoStop}%
\bibitem [{\citenamefont {Manohar}\ and\ \citenamefont
  {Wise}(1993)}]{Manohar:1992nd}%
  \BibitemOpen
  \bibfield  {author} {\bibinfo {author} {\bibfnamefont {A.~V.}\ \bibnamefont
  {Manohar}}\ and\ \bibinfo {author} {\bibfnamefont {M.~B.}\ \bibnamefont
  {Wise}},\ }\href {\doibase 10.1016/0550-3213(93)90614-U} {\bibfield
  {journal} {\bibinfo  {journal} {Nucl. Phys. B}\ }\textbf {\bibinfo {volume}
  {399}},\ \bibinfo {pages} {17} (\bibinfo {year} {1993})},\ \Eprint
  {http://arxiv.org/abs/hep-ph/9212236} {arXiv:hep-ph/9212236} \BibitemShut
  {NoStop}%
\bibitem [{\citenamefont {Meng}\ \emph {et~al.}(2021)\citenamefont {Meng},
  \citenamefont {Wang}, \citenamefont {Wang},\ and\ \citenamefont
  {Zhu}}]{Meng:2021jnw}%
  \BibitemOpen
  \bibfield  {author} {\bibinfo {author} {\bibfnamefont {L.}~\bibnamefont
  {Meng}}, \bibinfo {author} {\bibfnamefont {G.-J.}\ \bibnamefont {Wang}},
  \bibinfo {author} {\bibfnamefont {B.}~\bibnamefont {Wang}}, \ and\ \bibinfo
  {author} {\bibfnamefont {S.-L.}\ \bibnamefont {Zhu}},\ }\href@noop {} {\
  (\bibinfo {year} {2021})},\ \Eprint {http://arxiv.org/abs/2107.14784}
  {arXiv:2107.14784 [hep-ph]} \BibitemShut {NoStop}%
\bibitem [{\citenamefont {Agaev}\ \emph {et~al.}(2021)\citenamefont {Agaev},
  \citenamefont {Azizi},\ and\ \citenamefont {Sundu}}]{Agaev:2021vur}%
  \BibitemOpen
  \bibfield  {author} {\bibinfo {author} {\bibfnamefont {S.~S.}\ \bibnamefont
  {Agaev}}, \bibinfo {author} {\bibfnamefont {K.}~\bibnamefont {Azizi}}, \ and\
  \bibinfo {author} {\bibfnamefont {H.}~\bibnamefont {Sundu}},\ }\href@noop {}
  {\  (\bibinfo {year} {2021})},\ \Eprint {http://arxiv.org/abs/2108.00188}
  {arXiv:2108.00188 [hep-ph]} \BibitemShut {NoStop}%
\bibitem [{\citenamefont {Wu}\ \emph {et~al.}(2021{\natexlab{a}})\citenamefont
  {Wu}, \citenamefont {Pan}, \citenamefont {Liu}, \citenamefont {Luo},
  \citenamefont {Liu},\ and\ \citenamefont {Geng}}]{Wu:2021kbu}%
  \BibitemOpen
  \bibfield  {author} {\bibinfo {author} {\bibfnamefont {T.-W.}\ \bibnamefont
  {Wu}}, \bibinfo {author} {\bibfnamefont {Y.-W.}\ \bibnamefont {Pan}},
  \bibinfo {author} {\bibfnamefont {M.-Z.}\ \bibnamefont {Liu}}, \bibinfo
  {author} {\bibfnamefont {S.-Q.}\ \bibnamefont {Luo}}, \bibinfo {author}
  {\bibfnamefont {X.}~\bibnamefont {Liu}}, \ and\ \bibinfo {author}
  {\bibfnamefont {L.-S.}\ \bibnamefont {Geng}},\ }\href@noop {} {\  (\bibinfo
  {year} {2021}{\natexlab{a}})},\ \Eprint {http://arxiv.org/abs/2108.00923}
  {arXiv:2108.00923 [hep-ph]} \BibitemShut {NoStop}%
\bibitem [{\citenamefont {Ling}\ \emph {et~al.}(2021)\citenamefont {Ling},
  \citenamefont {Liu}, \citenamefont {Geng}, \citenamefont {Wang},\ and\
  \citenamefont {Xie}}]{Ling:2021bir}%
  \BibitemOpen
  \bibfield  {author} {\bibinfo {author} {\bibfnamefont {X.-Z.}\ \bibnamefont
  {Ling}}, \bibinfo {author} {\bibfnamefont {M.-Z.}\ \bibnamefont {Liu}},
  \bibinfo {author} {\bibfnamefont {L.-S.}\ \bibnamefont {Geng}}, \bibinfo
  {author} {\bibfnamefont {E.}~\bibnamefont {Wang}}, \ and\ \bibinfo {author}
  {\bibfnamefont {J.-J.}\ \bibnamefont {Xie}},\ }\href@noop {} {\  (\bibinfo
  {year} {2021})},\ \Eprint {http://arxiv.org/abs/2108.00947} {arXiv:2108.00947
  [hep-ph]} \BibitemShut {NoStop}%
\bibitem [{\citenamefont {Chen}\ \emph
  {et~al.}(2021{\natexlab{a}})\citenamefont {Chen}, \citenamefont {Huang},
  \citenamefont {Liu},\ and\ \citenamefont {Zhu}}]{Chen:2021vhg}%
  \BibitemOpen
  \bibfield  {author} {\bibinfo {author} {\bibfnamefont {R.}~\bibnamefont
  {Chen}}, \bibinfo {author} {\bibfnamefont {Q.}~\bibnamefont {Huang}},
  \bibinfo {author} {\bibfnamefont {X.}~\bibnamefont {Liu}}, \ and\ \bibinfo
  {author} {\bibfnamefont {S.-L.}\ \bibnamefont {Zhu}},\ }\href@noop {} {\
  (\bibinfo {year} {2021}{\natexlab{a}})},\ \Eprint
  {http://arxiv.org/abs/2108.01911} {arXiv:2108.01911 [hep-ph]} \BibitemShut
  {NoStop}%
\bibitem [{\citenamefont {Dong}\ \emph {et~al.}(2021)\citenamefont {Dong},
  \citenamefont {Guo},\ and\ \citenamefont {Zou}}]{Dong:2021bvy}%
  \BibitemOpen
  \bibfield  {author} {\bibinfo {author} {\bibfnamefont {X.-K.}\ \bibnamefont
  {Dong}}, \bibinfo {author} {\bibfnamefont {F.-K.}\ \bibnamefont {Guo}}, \
  and\ \bibinfo {author} {\bibfnamefont {B.-S.}\ \bibnamefont {Zou}},\
  }\href@noop {} {\  (\bibinfo {year} {2021})},\ \Eprint
  {http://arxiv.org/abs/2108.02673} {arXiv:2108.02673 [hep-ph]} \BibitemShut
  {NoStop}%
\bibitem [{\citenamefont {Feijoo}\ \emph {et~al.}(2021)\citenamefont {Feijoo},
  \citenamefont {Liang},\ and\ \citenamefont {Oset}}]{Feijoo:2021ppq}%
  \BibitemOpen
  \bibfield  {author} {\bibinfo {author} {\bibfnamefont {A.}~\bibnamefont
  {Feijoo}}, \bibinfo {author} {\bibfnamefont {W.~H.}\ \bibnamefont {Liang}}, \
  and\ \bibinfo {author} {\bibfnamefont {E.}~\bibnamefont {Oset}},\ }\href@noop
  {} {\  (\bibinfo {year} {2021})},\ \Eprint {http://arxiv.org/abs/2108.02730}
  {arXiv:2108.02730 [hep-ph]} \BibitemShut {NoStop}%
\bibitem [{\citenamefont {Yan}\ and\ \citenamefont
  {Valderrama}(2021)}]{Yan:2021wdl}%
  \BibitemOpen
  \bibfield  {author} {\bibinfo {author} {\bibfnamefont {M.-J.}\ \bibnamefont
  {Yan}}\ and\ \bibinfo {author} {\bibfnamefont {M.~P.}\ \bibnamefont
  {Valderrama}},\ }\href@noop {} {\  (\bibinfo {year} {2021})},\ \Eprint
  {http://arxiv.org/abs/2108.04785} {arXiv:2108.04785 [hep-ph]} \BibitemShut
  {NoStop}%
\bibitem [{\citenamefont {Dai}\ \emph {et~al.}(2021)\citenamefont {Dai},
  \citenamefont {Sun}, \citenamefont {Kang}, \citenamefont {Szczepaniak},\ and\
  \citenamefont {Yu}}]{Dai:2021wxi}%
  \BibitemOpen
  \bibfield  {author} {\bibinfo {author} {\bibfnamefont {L.-Y.}\ \bibnamefont
  {Dai}}, \bibinfo {author} {\bibfnamefont {X.}~\bibnamefont {Sun}}, \bibinfo
  {author} {\bibfnamefont {X.-W.}\ \bibnamefont {Kang}}, \bibinfo {author}
  {\bibfnamefont {A.~P.}\ \bibnamefont {Szczepaniak}}, \ and\ \bibinfo {author}
  {\bibfnamefont {J.-S.}\ \bibnamefont {Yu}},\ }\href@noop {} {\  (\bibinfo
  {year} {2021})},\ \Eprint {http://arxiv.org/abs/2108.06002} {arXiv:2108.06002
  [hep-ph]} \BibitemShut {NoStop}%
\bibitem [{\citenamefont {Weng}\ \emph {et~al.}(2021)\citenamefont {Weng},
  \citenamefont {Deng},\ and\ \citenamefont {Zhu}}]{Weng:2021hje}%
  \BibitemOpen
  \bibfield  {author} {\bibinfo {author} {\bibfnamefont {X.-Z.}\ \bibnamefont
  {Weng}}, \bibinfo {author} {\bibfnamefont {W.-Z.}\ \bibnamefont {Deng}}, \
  and\ \bibinfo {author} {\bibfnamefont {S.-L.}\ \bibnamefont {Zhu}},\
  }\href@noop {} {\  (\bibinfo {year} {2021})},\ \Eprint
  {http://arxiv.org/abs/2108.07242} {arXiv:2108.07242 [hep-ph]} \BibitemShut
  {NoStop}%
\bibitem [{\citenamefont {Huang}\ \emph {et~al.}(2021)\citenamefont {Huang},
  \citenamefont {Zhu}, \citenamefont {Geng},\ and\ \citenamefont
  {Wang}}]{Huang:2021urd}%
  \BibitemOpen
  \bibfield  {author} {\bibinfo {author} {\bibfnamefont {Y.}~\bibnamefont
  {Huang}}, \bibinfo {author} {\bibfnamefont {H.~Q.}\ \bibnamefont {Zhu}},
  \bibinfo {author} {\bibfnamefont {L.-S.}\ \bibnamefont {Geng}}, \ and\
  \bibinfo {author} {\bibfnamefont {R.}~\bibnamefont {Wang}},\ }\href@noop {}
  {\  (\bibinfo {year} {2021})},\ \Eprint {http://arxiv.org/abs/2108.13028}
  {arXiv:2108.13028 [hep-ph]} \BibitemShut {NoStop}%
\bibitem [{\citenamefont {Chen}\ \emph
  {et~al.}(2021{\natexlab{b}})\citenamefont {Chen}, \citenamefont {Li},
  \citenamefont {Sun}, \citenamefont {Liu},\ and\ \citenamefont
  {Zhu}}]{Chen:2021kad}%
  \BibitemOpen
  \bibfield  {author} {\bibinfo {author} {\bibfnamefont {R.}~\bibnamefont
  {Chen}}, \bibinfo {author} {\bibfnamefont {N.}~\bibnamefont {Li}}, \bibinfo
  {author} {\bibfnamefont {Z.-F.}\ \bibnamefont {Sun}}, \bibinfo {author}
  {\bibfnamefont {X.}~\bibnamefont {Liu}}, \ and\ \bibinfo {author}
  {\bibfnamefont {S.-L.}\ \bibnamefont {Zhu}},\ }\href@noop {} {\  (\bibinfo
  {year} {2021}{\natexlab{b}})},\ \Eprint {http://arxiv.org/abs/2108.12730}
  {arXiv:2108.12730 [hep-ph]} \BibitemShut {NoStop}%
\bibitem [{\citenamefont {Xin}\ and\ \citenamefont {Wang}(2021)}]{Xin:2021wcr}%
  \BibitemOpen
  \bibfield  {author} {\bibinfo {author} {\bibfnamefont {Q.}~\bibnamefont
  {Xin}}\ and\ \bibinfo {author} {\bibfnamefont {Z.-G.}\ \bibnamefont {Wang}},\
  }\href@noop {} {\  (\bibinfo {year} {2021})},\ \Eprint
  {http://arxiv.org/abs/2108.12597} {arXiv:2108.12597 [hep-ph]} \BibitemShut
  {NoStop}%
\bibitem [{\citenamefont {Hanhart}\ \emph {et~al.}(2015)\citenamefont
  {Hanhart}, \citenamefont {Kalashnikova}, \citenamefont {Matuschek},
  \citenamefont {Mizuk}, \citenamefont {Nefediev},\ and\ \citenamefont
  {Wang}}]{Hanhart_2015}%
  \BibitemOpen
  \bibfield  {author} {\bibinfo {author} {\bibfnamefont {C.}~\bibnamefont
  {Hanhart}}, \bibinfo {author} {\bibfnamefont {Y.}~\bibnamefont
  {Kalashnikova}}, \bibinfo {author} {\bibfnamefont {P.}~\bibnamefont
  {Matuschek}}, \bibinfo {author} {\bibfnamefont {R.}~\bibnamefont {Mizuk}},
  \bibinfo {author} {\bibfnamefont {A.}~\bibnamefont {Nefediev}}, \ and\
  \bibinfo {author} {\bibfnamefont {Q.}~\bibnamefont {Wang}},\ }\href {\doibase
  10.1103/physrevlett.115.202001} {\bibfield  {journal} {\bibinfo  {journal}
  {Physical Review Letters}\ }\textbf {\bibinfo {volume} {115}} (\bibinfo
  {year} {2015}),\ 10.1103/physrevlett.115.202001}\BibitemShut {NoStop}%
\bibitem [{\citenamefont {Aaij}\ \emph {et~al.}(2020)\citenamefont {Aaij} \emph
  {et~al.}}]{LHCb:2020xds}%
  \BibitemOpen
  \bibfield  {author} {\bibinfo {author} {\bibfnamefont {R.}~\bibnamefont
  {Aaij}} \emph {et~al.} (\bibinfo {collaboration} {LHCb}),\ }\href {\doibase
  10.1103/PhysRevD.102.092005} {\bibfield  {journal} {\bibinfo  {journal}
  {Phys. Rev. D}\ }\textbf {\bibinfo {volume} {102}},\ \bibinfo {pages}
  {092005} (\bibinfo {year} {2020})},\ \Eprint
  {http://arxiv.org/abs/2005.13419} {arXiv:2005.13419 [hep-ex]} \BibitemShut
  {NoStop}%
\bibitem [{\citenamefont {Mehen}(2015)}]{Mehen:2015efa}%
  \BibitemOpen
  \bibfield  {author} {\bibinfo {author} {\bibfnamefont {T.}~\bibnamefont
  {Mehen}},\ }\href {\doibase 10.1103/PhysRevD.92.034019} {\bibfield  {journal}
  {\bibinfo  {journal} {Phys. Rev.}\ }\textbf {\bibinfo {volume} {D92}},\
  \bibinfo {pages} {034019} (\bibinfo {year} {2015})},\ \Eprint
  {http://arxiv.org/abs/1503.02719} {arXiv:1503.02719 [hep-ph]} \BibitemShut
  {NoStop}%
\bibitem [{\citenamefont {Voloshin}(2004)}]{Voloshin:2003nt}%
  \BibitemOpen
  \bibfield  {author} {\bibinfo {author} {\bibfnamefont {M.~B.}\ \bibnamefont
  {Voloshin}},\ }\href {\doibase 10.1016/j.physletb.2003.11.014} {\bibfield
  {journal} {\bibinfo  {journal} {Phys. Lett.}\ }\textbf {\bibinfo {volume}
  {B579}},\ \bibinfo {pages} {316} (\bibinfo {year} {2004})},\ \Eprint
  {http://arxiv.org/abs/hep-ph/0309307} {arXiv:hep-ph/0309307 [hep-ph]}
  \BibitemShut {NoStop}%
\bibitem [{\citenamefont {Fleming}\ \emph {et~al.}(2007)\citenamefont
  {Fleming}, \citenamefont {Kusunoki}, \citenamefont {Mehen},\ and\
  \citenamefont {van Kolck}}]{Fleming:2007rp}%
  \BibitemOpen
  \bibfield  {author} {\bibinfo {author} {\bibfnamefont {S.}~\bibnamefont
  {Fleming}}, \bibinfo {author} {\bibfnamefont {M.}~\bibnamefont {Kusunoki}},
  \bibinfo {author} {\bibfnamefont {T.}~\bibnamefont {Mehen}}, \ and\ \bibinfo
  {author} {\bibfnamefont {U.}~\bibnamefont {van Kolck}},\ }\href {\doibase
  10.1103/PhysRevD.76.034006} {\bibfield  {journal} {\bibinfo  {journal} {Phys.
  Rev.}\ }\textbf {\bibinfo {volume} {D76}},\ \bibinfo {pages} {034006}
  (\bibinfo {year} {2007})},\ \Eprint {http://arxiv.org/abs/hep-ph/0703168}
  {arXiv:hep-ph/0703168 [hep-ph]} \BibitemShut {NoStop}%
\bibitem [{\citenamefont {Zyla}\ \emph {et~al.}(2020)\citenamefont {Zyla} \emph
  {et~al.}}]{Zyla:2020zbs}%
  \BibitemOpen
  \bibfield  {author} {\bibinfo {author} {\bibfnamefont {P.}~\bibnamefont
  {Zyla}} \emph {et~al.} (\bibinfo {collaboration} {Particle Data Group}),\
  }\href {\doibase 10.1093/ptep/ptaa104} {\bibfield  {journal} {\bibinfo
  {journal} {PTEP}\ }\textbf {\bibinfo {volume} {2020}},\ \bibinfo {pages}
  {083C01} (\bibinfo {year} {2020})}\BibitemShut {NoStop}%
\bibitem [{\citenamefont {Dai}\ \emph {et~al.}(2020)\citenamefont {Dai},
  \citenamefont {Guo},\ and\ \citenamefont {Mehen}}]{Dai:2019hrf}%
  \BibitemOpen
  \bibfield  {author} {\bibinfo {author} {\bibfnamefont {L.}~\bibnamefont
  {Dai}}, \bibinfo {author} {\bibfnamefont {F.-K.}\ \bibnamefont {Guo}}, \ and\
  \bibinfo {author} {\bibfnamefont {T.}~\bibnamefont {Mehen}},\ }\href
  {\doibase 10.1103/PhysRevD.101.054024} {\bibfield  {journal} {\bibinfo
  {journal} {Phys. Rev. D}\ }\textbf {\bibinfo {volume} {101}},\ \bibinfo
  {pages} {054024} (\bibinfo {year} {2020})},\ \Eprint
  {http://arxiv.org/abs/1912.04317} {arXiv:1912.04317 [hep-ph]} \BibitemShut
  {NoStop}%
\bibitem [{\citenamefont {Fleming}\ and\ \citenamefont
  {Mehen}(2008)}]{Fleming:2008yn}%
  \BibitemOpen
  \bibfield  {author} {\bibinfo {author} {\bibfnamefont {S.}~\bibnamefont
  {Fleming}}\ and\ \bibinfo {author} {\bibfnamefont {T.}~\bibnamefont
  {Mehen}},\ }\href {\doibase 10.1103/PhysRevD.78.094019} {\bibfield  {journal}
  {\bibinfo  {journal} {Phys. Rev.}\ }\textbf {\bibinfo {volume} {D78}},\
  \bibinfo {pages} {094019} (\bibinfo {year} {2008})},\ \Eprint
  {http://arxiv.org/abs/0807.2674} {arXiv:0807.2674 [hep-ph]} \BibitemShut
  {NoStop}%
\bibitem [{\citenamefont {Fleming}\ and\ \citenamefont
  {Mehen}(2012)}]{Fleming:2011xa}%
  \BibitemOpen
  \bibfield  {author} {\bibinfo {author} {\bibfnamefont {S.}~\bibnamefont
  {Fleming}}\ and\ \bibinfo {author} {\bibfnamefont {T.}~\bibnamefont
  {Mehen}},\ }\href {\doibase 10.1103/PhysRevD.85.014016} {\bibfield  {journal}
  {\bibinfo  {journal} {Phys. Rev.}\ }\textbf {\bibinfo {volume} {D85}},\
  \bibinfo {pages} {014016} (\bibinfo {year} {2012})},\ \Eprint
  {http://arxiv.org/abs/1110.0265} {arXiv:1110.0265 [hep-ph]} \BibitemShut
  {NoStop}%
\bibitem [{\citenamefont {Mehen}\ and\ \citenamefont
  {Springer}(2011)}]{Mehen:2011ds}%
  \BibitemOpen
  \bibfield  {author} {\bibinfo {author} {\bibfnamefont {T.}~\bibnamefont
  {Mehen}}\ and\ \bibinfo {author} {\bibfnamefont {R.}~\bibnamefont
  {Springer}},\ }\href {\doibase 10.1103/PhysRevD.83.094009} {\bibfield
  {journal} {\bibinfo  {journal} {Phys. Rev.}\ }\textbf {\bibinfo {volume}
  {D83}},\ \bibinfo {pages} {094009} (\bibinfo {year} {2011})},\ \Eprint
  {http://arxiv.org/abs/1101.5175} {arXiv:1101.5175 [hep-ph]} \BibitemShut
  {NoStop}%
\bibitem [{\citenamefont {Margaryan}\ and\ \citenamefont
  {Springer}(2013)}]{Margaryan:2013tta}%
  \BibitemOpen
  \bibfield  {author} {\bibinfo {author} {\bibfnamefont {A.}~\bibnamefont
  {Margaryan}}\ and\ \bibinfo {author} {\bibfnamefont {R.~P.}\ \bibnamefont
  {Springer}},\ }\href {\doibase 10.1103/PhysRevD.88.014017} {\bibfield
  {journal} {\bibinfo  {journal} {Phys. Rev.}\ }\textbf {\bibinfo {volume}
  {D88}},\ \bibinfo {pages} {014017} (\bibinfo {year} {2013})},\ \Eprint
  {http://arxiv.org/abs/1304.8101} {arXiv:1304.8101 [hep-ph]} \BibitemShut
  {NoStop}%
\bibitem [{\citenamefont {Braaten}\ \emph {et~al.}(2010)\citenamefont
  {Braaten}, \citenamefont {Hammer},\ and\ \citenamefont
  {Mehen}}]{Braaten:2010mg}%
  \BibitemOpen
  \bibfield  {author} {\bibinfo {author} {\bibfnamefont {E.}~\bibnamefont
  {Braaten}}, \bibinfo {author} {\bibfnamefont {H.-W.}\ \bibnamefont {Hammer}},
  \ and\ \bibinfo {author} {\bibfnamefont {T.}~\bibnamefont {Mehen}},\ }\href
  {\doibase 10.1103/PhysRevD.82.034018} {\bibfield  {journal} {\bibinfo
  {journal} {Phys. Rev.}\ }\textbf {\bibinfo {volume} {D82}},\ \bibinfo {pages}
  {034018} (\bibinfo {year} {2010})},\ \Eprint {http://arxiv.org/abs/1005.1688}
  {arXiv:1005.1688 [hep-ph]} \BibitemShut {NoStop}%
\bibitem [{\citenamefont {Canham}\ \emph {et~al.}(2009)\citenamefont {Canham},
  \citenamefont {Hammer},\ and\ \citenamefont {Springer}}]{Canham:2009zq}%
  \BibitemOpen
  \bibfield  {author} {\bibinfo {author} {\bibfnamefont {D.~L.}\ \bibnamefont
  {Canham}}, \bibinfo {author} {\bibfnamefont {H.-W.}\ \bibnamefont {Hammer}},
  \ and\ \bibinfo {author} {\bibfnamefont {R.~P.}\ \bibnamefont {Springer}},\
  }\href {\doibase 10.1103/PhysRevD.80.014009} {\bibfield  {journal} {\bibinfo
  {journal} {Phys. Rev.}\ }\textbf {\bibinfo {volume} {D80}},\ \bibinfo {pages}
  {014009} (\bibinfo {year} {2009})},\ \Eprint {http://arxiv.org/abs/0906.1263}
  {arXiv:0906.1263 [hep-ph]} \BibitemShut {NoStop}%
\bibitem [{\citenamefont {Jansen}\ \emph {et~al.}(2014)\citenamefont {Jansen},
  \citenamefont {Hammer},\ and\ \citenamefont {Jia}}]{Jansen:2013cba}%
  \BibitemOpen
  \bibfield  {author} {\bibinfo {author} {\bibfnamefont {M.}~\bibnamefont
  {Jansen}}, \bibinfo {author} {\bibfnamefont {H.-W.}\ \bibnamefont {Hammer}},
  \ and\ \bibinfo {author} {\bibfnamefont {Y.}~\bibnamefont {Jia}},\ }\href
  {\doibase 10.1103/PhysRevD.89.014033} {\bibfield  {journal} {\bibinfo
  {journal} {Phys. Rev.}\ }\textbf {\bibinfo {volume} {D89}},\ \bibinfo {pages}
  {014033} (\bibinfo {year} {2014})},\ \Eprint {http://arxiv.org/abs/1310.6937}
  {arXiv:1310.6937 [hep-ph]} \BibitemShut {NoStop}%
\bibitem [{\citenamefont {Jansen}\ \emph {et~al.}(2015)\citenamefont {Jansen},
  \citenamefont {Hammer},\ and\ \citenamefont {Jia}}]{Jansen:2015lha}%
  \BibitemOpen
  \bibfield  {author} {\bibinfo {author} {\bibfnamefont {M.}~\bibnamefont
  {Jansen}}, \bibinfo {author} {\bibfnamefont {H.-W.}\ \bibnamefont {Hammer}},
  \ and\ \bibinfo {author} {\bibfnamefont {Y.}~\bibnamefont {Jia}},\ }\href
  {\doibase 10.1103/PhysRevD.92.114031} {\bibfield  {journal} {\bibinfo
  {journal} {Phys. Rev.}\ }\textbf {\bibinfo {volume} {D92}},\ \bibinfo {pages}
  {114031} (\bibinfo {year} {2015})},\ \Eprint
  {http://arxiv.org/abs/1505.04099} {arXiv:1505.04099 [hep-ph]} \BibitemShut
  {NoStop}%
\bibitem [{\citenamefont {Alhakami}\ and\ \citenamefont
  {Birse}(2015)}]{Alhakami:2015uea}%
  \BibitemOpen
  \bibfield  {author} {\bibinfo {author} {\bibfnamefont {M.~H.}\ \bibnamefont
  {Alhakami}}\ and\ \bibinfo {author} {\bibfnamefont {M.~C.}\ \bibnamefont
  {Birse}},\ }\href {\doibase 10.1103/PhysRevD.91.054019} {\bibfield  {journal}
  {\bibinfo  {journal} {Phys. Rev.}\ }\textbf {\bibinfo {volume} {D91}},\
  \bibinfo {pages} {054019} (\bibinfo {year} {2015})},\ \Eprint
  {http://arxiv.org/abs/1501.06750} {arXiv:1501.06750 [hep-ph]} \BibitemShut
  {NoStop}%
\bibitem [{\citenamefont {Braaten}(2015)}]{Braaten:2015tga}%
  \BibitemOpen
  \bibfield  {author} {\bibinfo {author} {\bibfnamefont {E.}~\bibnamefont
  {Braaten}},\ }\href {\doibase 10.1103/PhysRevD.91.114007} {\bibfield
  {journal} {\bibinfo  {journal} {Phys. Rev.}\ }\textbf {\bibinfo {volume}
  {D91}},\ \bibinfo {pages} {114007} (\bibinfo {year} {2015})},\ \Eprint
  {http://arxiv.org/abs/1503.04791} {arXiv:1503.04791 [hep-ph]} \BibitemShut
  {NoStop}%
\bibitem [{\citenamefont {Braaten}\ \emph {et~al.}(2020)\citenamefont
  {Braaten}, \citenamefont {He}, \citenamefont {Ingles},\ and\ \citenamefont
  {Jiang}}]{Braaten:2020iye}%
  \BibitemOpen
  \bibfield  {author} {\bibinfo {author} {\bibfnamefont {E.}~\bibnamefont
  {Braaten}}, \bibinfo {author} {\bibfnamefont {L.-P.}\ \bibnamefont {He}},
  \bibinfo {author} {\bibfnamefont {K.}~\bibnamefont {Ingles}}, \ and\ \bibinfo
  {author} {\bibfnamefont {J.}~\bibnamefont {Jiang}},\ }\href {\doibase
  10.1103/PhysRevD.101.096020} {\bibfield  {journal} {\bibinfo  {journal}
  {Phys. Rev. D}\ }\textbf {\bibinfo {volume} {101}},\ \bibinfo {pages}
  {096020} (\bibinfo {year} {2020})},\ \Eprint
  {http://arxiv.org/abs/2004.12841} {arXiv:2004.12841 [hep-ph]} \BibitemShut
  {NoStop}%
\bibitem [{\citenamefont {Braaten}\ \emph
  {et~al.}(2021{\natexlab{a}})\citenamefont {Braaten}, \citenamefont {He},\
  and\ \citenamefont {Jiang}}]{Braaten:2020nmc}%
  \BibitemOpen
  \bibfield  {author} {\bibinfo {author} {\bibfnamefont {E.}~\bibnamefont
  {Braaten}}, \bibinfo {author} {\bibfnamefont {L.-P.}\ \bibnamefont {He}}, \
  and\ \bibinfo {author} {\bibfnamefont {J.}~\bibnamefont {Jiang}},\ }\href
  {\doibase 10.1103/PhysRevD.103.036014} {\bibfield  {journal} {\bibinfo
  {journal} {Phys. Rev. D}\ }\textbf {\bibinfo {volume} {103}},\ \bibinfo
  {pages} {036014} (\bibinfo {year} {2021}{\natexlab{a}})},\ \Eprint
  {http://arxiv.org/abs/2010.05801} {arXiv:2010.05801 [hep-ph]} \BibitemShut
  {NoStop}%
\bibitem [{\citenamefont {Braaten}\ \emph
  {et~al.}(2021{\natexlab{b}})\citenamefont {Braaten}, \citenamefont {He},
  \citenamefont {Ingles},\ and\ \citenamefont {Jiang}}]{Braaten:2020iqw}%
  \BibitemOpen
  \bibfield  {author} {\bibinfo {author} {\bibfnamefont {E.}~\bibnamefont
  {Braaten}}, \bibinfo {author} {\bibfnamefont {L.-P.}\ \bibnamefont {He}},
  \bibinfo {author} {\bibfnamefont {K.}~\bibnamefont {Ingles}}, \ and\ \bibinfo
  {author} {\bibfnamefont {J.}~\bibnamefont {Jiang}},\ }\href {\doibase
  10.1103/PhysRevD.103.L071901} {\bibfield  {journal} {\bibinfo  {journal}
  {Phys. Rev. D}\ }\textbf {\bibinfo {volume} {103}},\ \bibinfo {pages}
  {L071901} (\bibinfo {year} {2021}{\natexlab{b}})},\ \Eprint
  {http://arxiv.org/abs/2012.13499} {arXiv:2012.13499 [hep-ph]} \BibitemShut
  {NoStop}%
\bibitem [{\citenamefont {AlFiky}\ \emph {et~al.}(2006)\citenamefont {AlFiky},
  \citenamefont {Gabbiani},\ and\ \citenamefont {Petrov}}]{AlFiky:2005jd}%
  \BibitemOpen
  \bibfield  {author} {\bibinfo {author} {\bibfnamefont {M.~T.}\ \bibnamefont
  {AlFiky}}, \bibinfo {author} {\bibfnamefont {F.}~\bibnamefont {Gabbiani}}, \
  and\ \bibinfo {author} {\bibfnamefont {A.~A.}\ \bibnamefont {Petrov}},\
  }\href {\doibase 10.1016/j.physletb.2006.07.069} {\bibfield  {journal}
  {\bibinfo  {journal} {Phys. Lett.}\ }\textbf {\bibinfo {volume} {B640}},\
  \bibinfo {pages} {238} (\bibinfo {year} {2006})},\ \Eprint
  {http://arxiv.org/abs/hep-ph/0506141} {arXiv:hep-ph/0506141 [hep-ph]}
  \BibitemShut {NoStop}%
\bibitem [{\citenamefont {Baru}\ \emph {et~al.}(2011)\citenamefont {Baru},
  \citenamefont {Filin}, \citenamefont {Hanhart}, \citenamefont {Kalashnikova},
  \citenamefont {Kudryavtsev},\ and\ \citenamefont {Nefediev}}]{Baru:2011rs}%
  \BibitemOpen
  \bibfield  {author} {\bibinfo {author} {\bibfnamefont {V.}~\bibnamefont
  {Baru}}, \bibinfo {author} {\bibfnamefont {A.~A.}\ \bibnamefont {Filin}},
  \bibinfo {author} {\bibfnamefont {C.}~\bibnamefont {Hanhart}}, \bibinfo
  {author} {\bibfnamefont {{\relax Yu}.~S.}\ \bibnamefont {Kalashnikova}},
  \bibinfo {author} {\bibfnamefont {A.~E.}\ \bibnamefont {Kudryavtsev}}, \ and\
  \bibinfo {author} {\bibfnamefont {A.~V.}\ \bibnamefont {Nefediev}},\ }\href
  {\doibase 10.1103/PhysRevD.84.074029} {\bibfield  {journal} {\bibinfo
  {journal} {Phys. Rev.}\ }\textbf {\bibinfo {volume} {D84}},\ \bibinfo {pages}
  {074029} (\bibinfo {year} {2011})},\ \Eprint {http://arxiv.org/abs/1108.5644}
  {arXiv:1108.5644 [hep-ph]} \BibitemShut {NoStop}%
\bibitem [{\citenamefont {Valderrama}(2012)}]{Valderrama:2012jv}%
  \BibitemOpen
  \bibfield  {author} {\bibinfo {author} {\bibfnamefont {M.~P.}\ \bibnamefont
  {Valderrama}},\ }\href {\doibase 10.1103/PhysRevD.85.114037} {\bibfield
  {journal} {\bibinfo  {journal} {Phys. Rev.}\ }\textbf {\bibinfo {volume}
  {D85}},\ \bibinfo {pages} {114037} (\bibinfo {year} {2012})},\ \Eprint
  {http://arxiv.org/abs/1204.2400} {arXiv:1204.2400 [hep-ph]} \BibitemShut
  {NoStop}%
\bibitem [{\citenamefont {Nieves}\ and\ \citenamefont
  {Valderrama}(2012)}]{Nieves:2012tt}%
  \BibitemOpen
  \bibfield  {author} {\bibinfo {author} {\bibfnamefont {J.}~\bibnamefont
  {Nieves}}\ and\ \bibinfo {author} {\bibfnamefont {M.~P.}\ \bibnamefont
  {Valderrama}},\ }\href {\doibase 10.1103/PhysRevD.86.056004} {\bibfield
  {journal} {\bibinfo  {journal} {Phys. Rev.}\ }\textbf {\bibinfo {volume}
  {D86}},\ \bibinfo {pages} {056004} (\bibinfo {year} {2012})},\ \Eprint
  {http://arxiv.org/abs/1204.2790} {arXiv:1204.2790 [hep-ph]} \BibitemShut
  {NoStop}%
\bibitem [{\citenamefont {Baru}\ \emph {et~al.}(2013)\citenamefont {Baru},
  \citenamefont {Epelbaum}, \citenamefont {Filin}, \citenamefont {Hanhart},
  \citenamefont {Meissner},\ and\ \citenamefont {Nefediev}}]{Baru:2013rta}%
  \BibitemOpen
  \bibfield  {author} {\bibinfo {author} {\bibfnamefont {V.}~\bibnamefont
  {Baru}}, \bibinfo {author} {\bibfnamefont {E.}~\bibnamefont {Epelbaum}},
  \bibinfo {author} {\bibfnamefont {A.~A.}\ \bibnamefont {Filin}}, \bibinfo
  {author} {\bibfnamefont {C.}~\bibnamefont {Hanhart}}, \bibinfo {author}
  {\bibfnamefont {U.~G.}\ \bibnamefont {Meissner}}, \ and\ \bibinfo {author}
  {\bibfnamefont {A.~V.}\ \bibnamefont {Nefediev}},\ }\href {\doibase
  10.1016/j.physletb.2013.08.073} {\bibfield  {journal} {\bibinfo  {journal}
  {Phys. Lett.}\ }\textbf {\bibinfo {volume} {B726}},\ \bibinfo {pages} {537}
  (\bibinfo {year} {2013})},\ \Eprint {http://arxiv.org/abs/1306.4108}
  {arXiv:1306.4108 [hep-ph]} \BibitemShut {NoStop}%
\bibitem [{\citenamefont {Guo}\ \emph {et~al.}(2013{\natexlab{a}})\citenamefont
  {Guo}, \citenamefont {Hanhart}, \citenamefont {Mei{\ss}ner}, \citenamefont
  {Wang},\ and\ \citenamefont {Zhao}}]{Guo:2013nza}%
  \BibitemOpen
  \bibfield  {author} {\bibinfo {author} {\bibfnamefont {F.-K.}\ \bibnamefont
  {Guo}}, \bibinfo {author} {\bibfnamefont {C.}~\bibnamefont {Hanhart}},
  \bibinfo {author} {\bibfnamefont {U.-G.}\ \bibnamefont {Mei{\ss}ner}},
  \bibinfo {author} {\bibfnamefont {Q.}~\bibnamefont {Wang}}, \ and\ \bibinfo
  {author} {\bibfnamefont {Q.}~\bibnamefont {Zhao}},\ }\href {\doibase
  10.1016/j.physletb.2013.06.053} {\bibfield  {journal} {\bibinfo  {journal}
  {Phys. Lett.}\ }\textbf {\bibinfo {volume} {B725}},\ \bibinfo {pages} {127}
  (\bibinfo {year} {2013}{\natexlab{a}})},\ \Eprint
  {http://arxiv.org/abs/1306.3096} {arXiv:1306.3096 [hep-ph]} \BibitemShut
  {NoStop}%
\bibitem [{\citenamefont {Guo}\ \emph {et~al.}(2013{\natexlab{b}})\citenamefont
  {Guo}, \citenamefont {Hidalgo-Duque}, \citenamefont {Nieves},\ and\
  \citenamefont {Valderrama}}]{Guo:2013sya}%
  \BibitemOpen
  \bibfield  {author} {\bibinfo {author} {\bibfnamefont {F.-K.}\ \bibnamefont
  {Guo}}, \bibinfo {author} {\bibfnamefont {C.}~\bibnamefont {Hidalgo-Duque}},
  \bibinfo {author} {\bibfnamefont {J.}~\bibnamefont {Nieves}}, \ and\ \bibinfo
  {author} {\bibfnamefont {M.~P.}\ \bibnamefont {Valderrama}},\ }\href
  {\doibase 10.1103/PhysRevD.88.054007} {\bibfield  {journal} {\bibinfo
  {journal} {Phys. Rev.}\ }\textbf {\bibinfo {volume} {D88}},\ \bibinfo {pages}
  {054007} (\bibinfo {year} {2013}{\natexlab{b}})},\ \Eprint
  {http://arxiv.org/abs/1303.6608} {arXiv:1303.6608 [hep-ph]} \BibitemShut
  {NoStop}%
\bibitem [{\citenamefont {Baru}\ \emph {et~al.}(2015)\citenamefont {Baru},
  \citenamefont {Epelbaum}, \citenamefont {Filin}, \citenamefont {Guo},
  \citenamefont {Hammer}, \citenamefont {Hanhart}, \citenamefont {Mei§ner},\
  and\ \citenamefont {Nefediev}}]{Baru:2015nea}%
  \BibitemOpen
  \bibfield  {author} {\bibinfo {author} {\bibfnamefont {V.}~\bibnamefont
  {Baru}}, \bibinfo {author} {\bibfnamefont {E.}~\bibnamefont {Epelbaum}},
  \bibinfo {author} {\bibfnamefont {A.~A.}\ \bibnamefont {Filin}}, \bibinfo
  {author} {\bibfnamefont {F.-K.}\ \bibnamefont {Guo}}, \bibinfo {author}
  {\bibfnamefont {H.-W.}\ \bibnamefont {Hammer}}, \bibinfo {author}
  {\bibfnamefont {C.}~\bibnamefont {Hanhart}}, \bibinfo {author} {\bibfnamefont
  {U.-G.}\ \bibnamefont {Mei§ner}}, \ and\ \bibinfo {author} {\bibfnamefont
  {A.~V.}\ \bibnamefont {Nefediev}},\ }\href {\doibase
  10.1103/PhysRevD.91.034002} {\bibfield  {journal} {\bibinfo  {journal} {Phys.
  Rev.}\ }\textbf {\bibinfo {volume} {D91}},\ \bibinfo {pages} {034002}
  (\bibinfo {year} {2015})},\ \Eprint {http://arxiv.org/abs/1501.02924}
  {arXiv:1501.02924 [hep-ph]} \BibitemShut {NoStop}%
\bibitem [{\citenamefont {Schmidt}\ \emph {et~al.}(2018)\citenamefont
  {Schmidt}, \citenamefont {Jansen},\ and\ \citenamefont
  {Hammer}}]{Schmidt:2018vvl}%
  \BibitemOpen
  \bibfield  {author} {\bibinfo {author} {\bibfnamefont {M.}~\bibnamefont
  {Schmidt}}, \bibinfo {author} {\bibfnamefont {M.}~\bibnamefont {Jansen}}, \
  and\ \bibinfo {author} {\bibfnamefont {H.~W.}\ \bibnamefont {Hammer}},\
  }\href {\doibase 10.1103/PhysRevD.98.014032} {\bibfield  {journal} {\bibinfo
  {journal} {Phys. Rev.}\ }\textbf {\bibinfo {volume} {D98}},\ \bibinfo {pages}
  {014032} (\bibinfo {year} {2018})},\ \Eprint
  {http://arxiv.org/abs/1804.00375} {arXiv:1804.00375 [hep-ph]} \BibitemShut
  {NoStop}%
\bibitem [{\citenamefont {Sakai}\ \emph
  {et~al.}(2020{\natexlab{a}})\citenamefont {Sakai}, \citenamefont {Oset},\
  and\ \citenamefont {Guo}}]{Sakai:2020ucu}%
  \BibitemOpen
  \bibfield  {author} {\bibinfo {author} {\bibfnamefont {S.}~\bibnamefont
  {Sakai}}, \bibinfo {author} {\bibfnamefont {E.}~\bibnamefont {Oset}}, \ and\
  \bibinfo {author} {\bibfnamefont {F.-K.}\ \bibnamefont {Guo}},\ }\href
  {\doibase 10.1103/PhysRevD.101.054030} {\bibfield  {journal} {\bibinfo
  {journal} {Phys. Rev. D}\ }\textbf {\bibinfo {volume} {101}},\ \bibinfo
  {pages} {054030} (\bibinfo {year} {2020}{\natexlab{a}})},\ \Eprint
  {http://arxiv.org/abs/2002.03160} {arXiv:2002.03160 [hep-ph]} \BibitemShut
  {NoStop}%
\bibitem [{\citenamefont {Molina}\ and\ \citenamefont
  {Oset}(2020)}]{Molina:2020kyu}%
  \BibitemOpen
  \bibfield  {author} {\bibinfo {author} {\bibfnamefont {R.}~\bibnamefont
  {Molina}}\ and\ \bibinfo {author} {\bibfnamefont {E.}~\bibnamefont {Oset}},\
  }\href {\doibase 10.1140/epjc/s10052-020-8014-7} {\bibfield  {journal}
  {\bibinfo  {journal} {Eur. Phys. J. C}\ }\textbf {\bibinfo {volume} {80}},\
  \bibinfo {pages} {451} (\bibinfo {year} {2020})},\ \Eprint
  {http://arxiv.org/abs/2002.12821} {arXiv:2002.12821 [hep-ph]} \BibitemShut
  {NoStop}%
\bibitem [{\citenamefont {Sakai}\ \emph
  {et~al.}(2020{\natexlab{b}})\citenamefont {Sakai}, \citenamefont {Jing},\
  and\ \citenamefont {Guo}}]{Sakai:2020crh}%
  \BibitemOpen
  \bibfield  {author} {\bibinfo {author} {\bibfnamefont {S.}~\bibnamefont
  {Sakai}}, \bibinfo {author} {\bibfnamefont {H.-J.}\ \bibnamefont {Jing}}, \
  and\ \bibinfo {author} {\bibfnamefont {F.-K.}\ \bibnamefont {Guo}},\ }\href
  {\doibase 10.1103/PhysRevD.102.114041} {\bibfield  {journal} {\bibinfo
  {journal} {Phys. Rev. D}\ }\textbf {\bibinfo {volume} {102}},\ \bibinfo
  {pages} {114041} (\bibinfo {year} {2020}{\natexlab{b}})},\ \Eprint
  {http://arxiv.org/abs/2008.10829} {arXiv:2008.10829 [hep-ph]} \BibitemShut
  {NoStop}%
\bibitem [{\citenamefont {Contessi}\ \emph {et~al.}(2021)\citenamefont
  {Contessi}, \citenamefont {Kirscher},\ and\ \citenamefont
  {Pavon~Valderrama}}]{Contessi:2020jqa}%
  \BibitemOpen
  \bibfield  {author} {\bibinfo {author} {\bibfnamefont {L.}~\bibnamefont
  {Contessi}}, \bibinfo {author} {\bibfnamefont {J.}~\bibnamefont {Kirscher}},
  \ and\ \bibinfo {author} {\bibfnamefont {M.}~\bibnamefont
  {Pavon~Valderrama}},\ }\href {\doibase 10.1103/PhysRevD.103.056001}
  {\bibfield  {journal} {\bibinfo  {journal} {Phys. Rev. D}\ }\textbf {\bibinfo
  {volume} {103}},\ \bibinfo {pages} {056001} (\bibinfo {year} {2021})},\
  \Eprint {http://arxiv.org/abs/2008.12268} {arXiv:2008.12268 [hep-ph]}
  \BibitemShut {NoStop}%
\bibitem [{\citenamefont {Wu}\ \emph {et~al.}(2021{\natexlab{b}})\citenamefont
  {Wu}, \citenamefont {Chen},\ and\ \citenamefont {Matsuki}}]{Wu:2021udi}%
  \BibitemOpen
  \bibfield  {author} {\bibinfo {author} {\bibfnamefont {Q.}~\bibnamefont
  {Wu}}, \bibinfo {author} {\bibfnamefont {D.-Y.}\ \bibnamefont {Chen}}, \ and\
  \bibinfo {author} {\bibfnamefont {T.}~\bibnamefont {Matsuki}},\ }\href
  {\doibase 10.1140/epjc/s10052-021-08984-2} {\bibfield  {journal} {\bibinfo
  {journal} {Eur. Phys. J. C}\ }\textbf {\bibinfo {volume} {81}},\ \bibinfo
  {pages} {193} (\bibinfo {year} {2021}{\natexlab{b}})},\ \Eprint
  {http://arxiv.org/abs/2102.08637} {arXiv:2102.08637 [hep-ph]} \BibitemShut
  {NoStop}%
\bibitem [{\citenamefont {Wise}(1992)}]{Wise:1992hn}%
  \BibitemOpen
  \bibfield  {author} {\bibinfo {author} {\bibfnamefont {M.~B.}\ \bibnamefont
  {Wise}},\ }\href {\doibase 10.1103/PhysRevD.45.R2188} {\bibfield  {journal}
  {\bibinfo  {journal} {Phys. Rev.}\ }\textbf {\bibinfo {volume} {D45}},\
  \bibinfo {pages} {R2188} (\bibinfo {year} {1992})}\BibitemShut {NoStop}%
\bibitem [{\citenamefont {Burdman}\ and\ \citenamefont
  {Donoghue}(1992)}]{Burdman:1992gh}%
  \BibitemOpen
  \bibfield  {author} {\bibinfo {author} {\bibfnamefont {G.}~\bibnamefont
  {Burdman}}\ and\ \bibinfo {author} {\bibfnamefont {J.~F.}\ \bibnamefont
  {Donoghue}},\ }\href {\doibase 10.1016/0370-2693(92)90068-F} {\bibfield
  {journal} {\bibinfo  {journal} {Phys. Lett.}\ }\textbf {\bibinfo {volume}
  {B280}},\ \bibinfo {pages} {287} (\bibinfo {year} {1992})}\BibitemShut
  {NoStop}%
\bibitem [{\citenamefont {Yan}\ \emph {et~al.}(1992)\citenamefont {Yan},
  \citenamefont {Cheng}, \citenamefont {Cheung}, \citenamefont {Lin},
  \citenamefont {Lin},\ and\ \citenamefont {Yu}}]{Yan:1992gz}%
  \BibitemOpen
  \bibfield  {author} {\bibinfo {author} {\bibfnamefont {T.-M.}\ \bibnamefont
  {Yan}}, \bibinfo {author} {\bibfnamefont {H.-Y.}\ \bibnamefont {Cheng}},
  \bibinfo {author} {\bibfnamefont {C.-Y.}\ \bibnamefont {Cheung}}, \bibinfo
  {author} {\bibfnamefont {G.-L.}\ \bibnamefont {Lin}}, \bibinfo {author}
  {\bibfnamefont {Y.~C.}\ \bibnamefont {Lin}}, \ and\ \bibinfo {author}
  {\bibfnamefont {H.-L.}\ \bibnamefont {Yu}},\ }\href {\doibase
  10.1103/PhysRevD.46.1148, 10.1103/PhysRevD.55.5851} {\bibfield  {journal}
  {\bibinfo  {journal} {Phys. Rev.}\ }\textbf {\bibinfo {volume} {D46}},\
  \bibinfo {pages} {1148} (\bibinfo {year} {1992})},\ \bibinfo {note}
  {[Erratum: Phys. Rev.D55,5851(1997)]}\BibitemShut {NoStop}%
\bibitem [{\citenamefont {Amundson}\ \emph {et~al.}(1992)\citenamefont
  {Amundson}, \citenamefont {Boyd}, \citenamefont {Jenkins}, \citenamefont
  {Luke}, \citenamefont {Manohar}, \citenamefont {Rosner}, \citenamefont
  {Savage},\ and\ \citenamefont {Wise}}]{Amundson:1992yp}%
  \BibitemOpen
  \bibfield  {author} {\bibinfo {author} {\bibfnamefont {J.~F.}\ \bibnamefont
  {Amundson}}, \bibinfo {author} {\bibfnamefont {C.~G.}\ \bibnamefont {Boyd}},
  \bibinfo {author} {\bibfnamefont {E.~E.}\ \bibnamefont {Jenkins}}, \bibinfo
  {author} {\bibfnamefont {M.~E.}\ \bibnamefont {Luke}}, \bibinfo {author}
  {\bibfnamefont {A.~V.}\ \bibnamefont {Manohar}}, \bibinfo {author}
  {\bibfnamefont {J.~L.}\ \bibnamefont {Rosner}}, \bibinfo {author}
  {\bibfnamefont {M.~J.}\ \bibnamefont {Savage}}, \ and\ \bibinfo {author}
  {\bibfnamefont {M.~B.}\ \bibnamefont {Wise}},\ }\href {\doibase
  10.1016/0370-2693(92)91341-6} {\bibfield  {journal} {\bibinfo  {journal}
  {Phys. Lett. B}\ }\textbf {\bibinfo {volume} {296}},\ \bibinfo {pages} {415}
  (\bibinfo {year} {1992})},\ \Eprint {http://arxiv.org/abs/hep-ph/9209241}
  {arXiv:hep-ph/9209241} \BibitemShut {NoStop}%
\bibitem [{\citenamefont {Stewart}(1998)}]{Stewart:1998ke}%
  \BibitemOpen
  \bibfield  {author} {\bibinfo {author} {\bibfnamefont {I.~W.}\ \bibnamefont
  {Stewart}},\ }\href {\doibase 10.1016/S0550-3213(98)00374-5} {\bibfield
  {journal} {\bibinfo  {journal} {Nucl. Phys. B}\ }\textbf {\bibinfo {volume}
  {529}},\ \bibinfo {pages} {62} (\bibinfo {year} {1998})},\ \Eprint
  {http://arxiv.org/abs/hep-ph/9803227} {arXiv:hep-ph/9803227} \BibitemShut
  {NoStop}%
\bibitem [{\citenamefont {Hu}\ and\ \citenamefont {Mehen}(2006)}]{Hu:2005gf}%
  \BibitemOpen
  \bibfield  {author} {\bibinfo {author} {\bibfnamefont {J.}~\bibnamefont
  {Hu}}\ and\ \bibinfo {author} {\bibfnamefont {T.}~\bibnamefont {Mehen}},\
  }\href {\doibase 10.1103/PhysRevD.73.054003} {\bibfield  {journal} {\bibinfo
  {journal} {Phys. Rev. D}\ }\textbf {\bibinfo {volume} {73}},\ \bibinfo
  {pages} {054003} (\bibinfo {year} {2006})},\ \Eprint
  {http://arxiv.org/abs/hep-ph/0511321} {arXiv:hep-ph/0511321} \BibitemShut
  {NoStop}%
\bibitem [{\citenamefont {Mehen}\ and\ \citenamefont
  {Powell}(2011)}]{Mehen:2011yh}%
  \BibitemOpen
  \bibfield  {author} {\bibinfo {author} {\bibfnamefont {T.}~\bibnamefont
  {Mehen}}\ and\ \bibinfo {author} {\bibfnamefont {J.~W.}\ \bibnamefont
  {Powell}},\ }\href {\doibase 10.1103/PhysRevD.84.114013} {\bibfield
  {journal} {\bibinfo  {journal} {Phys. Rev.}\ }\textbf {\bibinfo {volume}
  {D84}},\ \bibinfo {pages} {114013} (\bibinfo {year} {2011})},\ \Eprint
  {http://arxiv.org/abs/1109.3479} {arXiv:1109.3479 [hep-ph]} \BibitemShut
  {NoStop}%
\bibitem [{\citenamefont {Liu}\ \emph {et~al.}(2013)\citenamefont {Liu},
  \citenamefont {Orginos}, \citenamefont {Guo}, \citenamefont {Hanhart},\ and\
  \citenamefont {Mei{\ss}ner}}]{Liu:2012zya}%
  \BibitemOpen
  \bibfield  {author} {\bibinfo {author} {\bibfnamefont {L.}~\bibnamefont
  {Liu}}, \bibinfo {author} {\bibfnamefont {K.}~\bibnamefont {Orginos}},
  \bibinfo {author} {\bibfnamefont {F.-K.}\ \bibnamefont {Guo}}, \bibinfo
  {author} {\bibfnamefont {C.}~\bibnamefont {Hanhart}}, \ and\ \bibinfo
  {author} {\bibfnamefont {U.-G.}\ \bibnamefont {Mei{\ss}ner}},\ }\href
  {\doibase 10.1103/PhysRevD.87.014508} {\bibfield  {journal} {\bibinfo
  {journal} {Phys. Rev.}\ }\textbf {\bibinfo {volume} {D87}},\ \bibinfo {pages}
  {014508} (\bibinfo {year} {2013})},\ \Eprint {http://arxiv.org/abs/1208.4535}
  {arXiv:1208.4535 [hep-lat]} \BibitemShut {NoStop}%
\bibitem [{\citenamefont {Mohler}\ \emph {et~al.}(2013)\citenamefont {Mohler},
  \citenamefont {Prelovsek},\ and\ \citenamefont {Woloshyn}}]{Mohler_2013}%
  \BibitemOpen
  \bibfield  {author} {\bibinfo {author} {\bibfnamefont {D.}~\bibnamefont
  {Mohler}}, \bibinfo {author} {\bibfnamefont {S.}~\bibnamefont {Prelovsek}}, \
  and\ \bibinfo {author} {\bibfnamefont {R.~M.}\ \bibnamefont {Woloshyn}},\
  }\href {\doibase 10.1103/physrevd.87.034501} {\bibfield  {journal} {\bibinfo
  {journal} {Physical Review D}\ }\textbf {\bibinfo {volume} {87}} (\bibinfo
  {year} {2013}),\ 10.1103/physrevd.87.034501}\BibitemShut {NoStop}%
\bibitem [{\citenamefont {Tan}\ \emph {et~al.}(2020)\citenamefont {Tan},
  \citenamefont {Lu},\ and\ \citenamefont {Ping}}]{Tan_2020}%
  \BibitemOpen
  \bibfield  {author} {\bibinfo {author} {\bibfnamefont {Y.}~\bibnamefont
  {Tan}}, \bibinfo {author} {\bibfnamefont {W.}~\bibnamefont {Lu}}, \ and\
  \bibinfo {author} {\bibfnamefont {J.}~\bibnamefont {Ping}},\ }\href {\doibase
  10.1140/epjp/s13360-020-00741-w} {\bibfield  {journal} {\bibinfo  {journal}
  {The European Physical Journal Plus}\ }\textbf {\bibinfo {volume} {135}}
  (\bibinfo {year} {2020}),\ 10.1140/epjp/s13360-020-00741-w}\BibitemShut
  {NoStop}%
\bibitem [{\citenamefont {Yang}\ \emph {et~al.}(2009)\citenamefont {Yang},
  \citenamefont {Deng}, \citenamefont {Ping},\ and\ \citenamefont
  {Goldman}}]{PhysRevD.80.114023}%
  \BibitemOpen
  \bibfield  {author} {\bibinfo {author} {\bibfnamefont {Y.}~\bibnamefont
  {Yang}}, \bibinfo {author} {\bibfnamefont {C.}~\bibnamefont {Deng}}, \bibinfo
  {author} {\bibfnamefont {J.}~\bibnamefont {Ping}}, \ and\ \bibinfo {author}
  {\bibfnamefont {T.}~\bibnamefont {Goldman}},\ }\href {\doibase
  10.1103/PhysRevD.80.114023} {\bibfield  {journal} {\bibinfo  {journal} {Phys.
  Rev. D}\ }\textbf {\bibinfo {volume} {80}},\ \bibinfo {pages} {114023}
  (\bibinfo {year} {2009})}\BibitemShut {NoStop}%
\end{thebibliography}%

\end{document}